\documentclass[aps,prl,twocolumn,showpacs,floatfix,longbibliography,superscriptaddress]{revtex4-1}
\usepackage{graphicx,amsfonts,amssymb,amsmath}
\usepackage{textcomp}
\usepackage{esint}
\usepackage[colorlinks,linktocpage,bookmarks=false,citecolor=blue,linkcolor=red,urlcolor=blue,hypertexnames=false]{hyperref}
\usepackage[T1]{fontenc}
\usepackage[dvipsnames]{xcolor}
\usepackage[english]{babel}
\usepackage{afterpage}
\usepackage{soul}
\usepackage{subfigure} 

\allowdisplaybreaks

\def\cthree{C_{\rm 3D}}

\def\cquasi{C}
\def\Pquasi{P}
\def\Oquasi{\Omega}
\def\Eqd{E}

\begin{document}


\def\rhoeq{\hat\rho_{\rm eq}}

\newcommand{\marge}[1]{\marginpar{\scriptsize #1}}
\newcommand{\remarque}[1]{\marginpar{\scriptsize Remarque}{\it [#1]}}
\newcommand{\new}[1]{{\bf #1}}
\newcommand{\red}[1]{\textcolor{red}{#1}}
\newcommand{\blue}[1]{\textcolor{blue}{#1}}
\newcommand{\green}[1]{\textcolor{green}{#1}}
\newlength{\textlarg}
\newcommand{\redbar}[1]{\textcolor{red}{\st{#1}}} 
\newcommand{\bluebar}[1]{\textcolor{blue}{\st{#1}}} 

\newcommand{\beq}{\begin{equation}}
\newcommand{\eeq}{\end{equation}}
\newcommand{\bfig}{\begin{figure}}
\newcommand{\efig}{\end{figure}}
\newcommand{\bline}{\begin{multline}}
\newcommand{\eline}{\end{multline}}
\newcommand{\bremark}{\begin{quotation} \noindent \small }
\newcommand{\eremark}{\end{quotation}}
\newcommand{\llbrace}{\left\lbrace}  
\newcommand{\rrbrace}{\right\rbrace}
\newcommand{\lbraket}{\left[}
\newcommand{\rbraket}{\right]}
\newcommand{\llangle}{\left\langle}
\newcommand{\rrangle}{\right\rangle} 

\newcommand{\Tr}{{\rm Tr}} 
\newcommand{\tr}{{\rm tr}} 
\newcommand{\sgn}{\,{\rm sgn}} 
\newcommand{\mean}[1]{\langle #1 \rangle}
\newcommand{\commu}[2]{[#1,#2]} 
\newcommand{\bra}[1]{\langle#1|}
\newcommand{\ket}[1]{|#1\rangle}
\newcommand{\braket}[2]{\langle #1|#2\rangle}
\newcommand{\ketbra}[2]{|#1\rangle\langle#2|}
\newcommand{\dbraket}[3]{\langle #1|#2|#3\rangle}
\newcommand{\tens}[1]{\overleftrightarrow{#1}}  
\newcommand{\vac}{|{\rm vac}\rangle} 
\newcommand{\bravac}{\langle{\rm vac}|}
\newcommand{\const}{{\rm const}} 
\newcommand{\unif}{{\rm unif.}} 
\newcommand{\atanh}{\,{\rm atanh}}
\newcommand{\cotanh}{\,{\rm cotanh}}

\newcommand{\ie}{i.e.\xspace}
\newcommand{\iet}{i.e.}
\newcommand{\eg}{e.g.\xspace}
\newcommand{\cc}{{\rm c.c.}} 
\newcommand{\hc}{{\rm h.c.}} 
\newcommand{\etal}{{\it et al. }}
\newcommand\eme{$^{\mbox{\footnotesize ème}}$\xspace}

\newcommand{\jhatbf}{\hat {\textbf \jold}} 
\newcommand{\Jhatbf}{\hat {\textbf \J}} 
\newcommand{\jhat}{\hat {\jmath}} 
\newcommand{\Jhat}{\hat {J}} 
\newcommand{\jbf}{\textbf j}
\newcommand{\Jbf}{\textbf J}

\def\chibf{\boldsymbol{\chi}}
\def\down{\downarrow}
\def\eps{\epsilon}
\def\gam{\gamma} 
\def\alphabf{\boldsymbol{\alpha}}
\def\phibf{\boldsymbol{\phi}}
\def\varphibf{\boldsymbol{\varphi}}
\def\varphibfs{\boldsymbol{\varphi}_<}
\def\varphibfl{\boldsymbol{\varphi}_>}
\def\varphis{\varphi_{<}}
\def\varphil{\varphi_{>}}
\def\psibf{\boldsymbol{\psi}}
\def\thetabf{\boldsymbol{\theta}}
\def\Ome{\Omega}
\def\omeD{{\omega_D}} 
\def\bfOme{\boldsymbol{\Omega}} 
\def\Omebf{\boldsymbol{\Omega}} 
\def\lamb{\lambda}
\def\Lamb{\Lambda}
\def\sig{\sigma}
\def\Sig{\Sigma}
\def\sigp{{\sigma'}} 
\def\bfsig{\boldsymbol{\sigma}} 
\def\sigbf{\boldsymbol{\sigma}} 
\def\bfSig{\boldsymbol{\Sigma}} 
\def\The{\Theta} 
\def\up{\uparrow}

\def\epsk{\epsilon_{\bf k}} 
\def\xik{\xi_{\bf k}} 
\def\txik{\tilde\xi_{\bf k}} 
\def\xip{\xi_{\bf p}} 
\def\epsq{\epsilon_{\bf q}} 
\def\xiq{\xi_{\bf q}} 
\def\xikq{\xi_{{\bf k}+{\bf q}}} 
\def\Ek{E_{\bf k}} 
\def\Ep{E_{\bf p}}
\def\Eq{E_{\bf q}}
\def\Heff{\hat H_{\rm eff}}
\def\Hem{\hat H_{\rm em}}
\def\Hint{\hat H_{\rm int}}
\def\Hloc{\hat H_{\rm loc}}
\def\HMF{\hat H_{\rm MF}}
\def\HLL{\hat H_{\rm LL}}
\def\Sem{S_{\rm em}}
\def\SMF{S_{\rm MF}} 
\def\SHF{S_{\rm HF}} 
\def\SRPA{S_{\rm RPA}} 
\def\Sint{S_{\rm int}} 
\def\Sloc{S_{\rm loc}}
\def\TN{T_{\rm N}} 
\def\TNHF{T^{\rm HF}_{\rm N}} 
\def\Zloc{Z_{\rm loc}} 
\def\ZMF{Z_{\rm MF}} 
\def\ZHF{Z_{\rm HF}} 
\def\ZRPA{Z_{\rm RPA}} 
\def\RPA{{\rm RPA}}
\def\loc{{\rm loc}} 
\def\pp{{\rm pp}}
\def\ph{{\rm ph}} 
\def\ch{{\rm ch}}
\def\sp{{\rm sp}} 
\def\qtf{q_{\rm TF}}
\def\epstf{\eps^{}_{\rm TF}} 
\def\epsrpa{\eps^{}_{\rm RPA}} 
\def\chinnzpp{\chi_{nn}^{0}{}\!\!\!''}

\def\half{\frac{1}{2}}
\def\dhalf{\dfrac{1}{2}}
\def\third{\frac{1}{3}} 
\def\quarter{\frac{1}{4}}

\def\qr{{\bf q}\cdot{\bf r}}
\def\wt{\omega t} 

\def\a{{\bf a}}
\def\b{{\bf b}}
\newcommand{\cv}{{\bf c}} 
\def\e{{\bf e}}
\def\f{{\bf f}}
\def\g{{\bf g}}
\def\h{{\bf h}}
\def\jold{\char"11}
\def\j{{\bf j}}
\def\k{{\bf k}}
\def\l{{\bf l}}
\def\ellbf{\bm{\ell}} 
\def\m{{\bf m}}
\def\n{{\bf n}} 
\def\p{{\bf p}} 
\def\q{{\bf q}}
\def\r{{\bf r}}
\def\t{{\bf t}}
\def\u{{\bf u}}
\newcommand{\vv}{{\bf v}}
\def\x{{\bf x}}
\def\y{{\bf y}} 
\def\z{{\bf z}} 
\def\A{{\bf A}}
\def\B{{\bf B}}
\def\D{{\bf D}} 
\def\E{{\bf E}} 
\def\F{{\bf F}} 
\def\H{{\bf H}}  
\def\J{{\bf J}}
\def\K{{\bf K}} 

\def\G{{\bf G}}
\def\L{{\bf L}}
\def\M{{\bf M}}  
\def\O{{\bf O}} 
\def\P{{\bf P}} 
\def\Q{{\bf Q}} 
\def\R{{\bf R}}
\def\S{{\bf S}}
\def\U{{\bf U}} 
\def\V{{\bf V}} 
\def\X{{\bf X}} 
\def\Y{{\bf Y}} 
\def\epsbf{\boldsymbol{\epsilon}}
\def\betabf{\boldsymbol{\beta}}
\def\deltabf{\boldsymbol{\delta}}
\def\mubf{\boldsymbol{\mu}}
\def\nablabf{\boldsymbol{\nabla}}
\def\rhobf{\boldsymbol{\rho}}
\def\sigmabf{\boldsymbol{\sigma}} 
\def\Pibf{\boldsymbol{\Pi}}
\def\pibf{\boldsymbol{\pi}}

\def\para{\parallel}
\def\kpara{{k_\parallel}}
\def\kperp{{k_\perp}} 
\def\kperpp{{k_\perp'}} 
\def\qperp{{q_\perp}} 
\def\tperp{{t_\perp}} 

\def\w{\omega}
\def\wn{\omega_n}
\def\wm{\omega_m}
\def\wnu{\omega_\nu}
\def\wp{\omega_p} 
\def\dmu{{\partial_\mu}}
\def\dnu{{\partial_\nu}}
\def\dl{{\partial_l}}  
\def\dt{\partial_t} 
\def\tdt{\tilde\partial_t}
\def\dk{\partial_k}
\def\tdk{\tilde\partial_k}
\def\dx{\partial_x}
\def\dy{\partial_y} 
\def\dw{\partial_{\w}}
\def\dtau{{\partial_\tau}}  
\def\det{{\rm det}} 
\def\Pf{{\rm Pf}}
\def\diag{{\rm diag}}

\def\dsum{\displaystyle \sum}
\def\dint{\displaystyle \int} 
\def\intt{\int_{-\infty}^\infty dt} 
\def\inttp{\int_{-\infty}^\infty dt'} 
\def\intk{\int_{\bf k}} 
\def\intkd{\int \frac{d^dk}{(2\pi)^d}}
\def\intq{\int_{\bf q}} 
\def\intr{\int d^dr}  
\def\dintr{\displaystyle \int d^dr} 
\def\intrp{\int d^dr'}
\def\dinttau{\displaystyle \int_0^\beta d\tau}
\def\dinttaup{\displaystyle \int_0^\beta d\tau'}
\def\inttau{\int_0^\beta d\tau}
\def\inttaup{\int_0^\beta d\tau'}
\def\intx{\int d^{d+1}x} 
\def\inttaur{\int_0^\beta d\tau \int d^dr}
\def\intinf{\int_{-\infty}^\infty}
\def\dinttaur{\displaystyle \int_0^\beta d\tau \int d^dr}
\def\dintinf{\displaystyle \int_{-\infty}^\infty}
\def\intw{\int_{-\infty}^\infty \frac{d\w}{2\pi}}
\def\sumr{\sum_{\bf r}} 

\def\calA{{\cal A}}
\def\calAbf{\bm{{\cal A}}}
\def\calB{{\cal B}} 
\def\calC{{\cal C}} 
\def\dt{\partial_t}
\def\calD{{\cal D}}
\def\calE{{\cal E}}
\def\calF{{\cal F}} 
\def\calFbf{\bm{{\cal F}}}
\def\calG{{\cal G}}
\def\calH{{\cal H}}
\def\calI{{\cal I}}
\def\calJ{{\cal J}}
\def\calK{{\cal K}}
\def\calL{{\cal L}} 
\def\calM{{\cal M}} 
\def\calN{{\cal N}}
\def\calO{{\cal O}}
\def\calP{{\cal P}}  
\def\calR{{\cal R}} 
\def\calS{{\cal S}}
\def\calT{{\cal T}}
\def\calU{{\cal U}}
\def\calV{{\cal V}}
\def\calX{{\cal X}} 
\def\calY{{\cal Y}} 
\def\calW{{\cal W}} 
\def\calZ{{\cal Z}}

\def\tT{{\tilde T}}
\def\talpha{{\tilde\alpha}}
\def\tbeta{{\tilde\beta}}
\def\tchi{{\tilde\chi}}
\def\tdelta{{\tilde\delta}}
\def\tDelta{{\tilde\Delta}}
\def\teta{{\tilde\eta}} 
\def\tlamb{{\tilde\lambda}}
\def\tmu{{\tilde\mu}}
\def\tphibf{{\tilde\phibf}}
\def\trho{{\tilde\rho}}
\def\tvarphibf{{\tilde\varphibf}} 
\def\tw{{\tilde\omega}}
\def\twn{{\tilde\omega_n}}
\def\twnu{{\tilde\omega_\nu}}

\def\asinh{{\rm asinh}} 
\def\Tbkt{T_{\rm BKT}}

\title{Tan's two-body contact in a planar Bose gas: experiment {\it vs} theory} 
\author{Adam Ran\c{c}on}
\affiliation{Univ. Lille, CNRS, UMR 8523 – PhLAM – Laboratoire de Physique des Lasers Atomes et Molécules, F-59000 Lille, France}
\affiliation{Institute  of  Physics,  Bijeni\v cka  cesta  46,  HR-10001  Zagreb,  Croatia}
\author{Nicolas Dupuis}
\affiliation{Sorbonne Universit\'e, CNRS, Laboratoire de Physique Th\'eorique de la Mati\`ere Condens\'ee, LPTMC, F-75005 Paris, France}

\date{June 20, 2023} 

\begin{abstract}
	We determine the two-body contact in a planar Bose gas confined by a transverse harmonic potential, using the nonperturbative functional renormalization group.
	We use the three-dimensional thermodynamic definition of the contact where the latter is related to the derivation of the pressure of the quasi-two-dimensional system with respect to the three-dimensional scattering length of the bosons. Without any free parameter, we find a remarkable agreement with the experimental data of Zou {\it et al.} [Nat. Comm. {\bf 12}, 760 (2021)] from low to high temperatures, including the vicinity of the Berezinskii-Kosterlitz-Thouless transition. We also show that the short-distance behavior of the pair distribution function and the high-momentum behavior of the momentum distribution are determined by two contacts: the three-dimensional contact for length scales smaller than the characteristic length $\ell_z=\sqrt{\hbar/m\omega_z}$ of the harmonic potential and, for length scales larger than $\ell_z$, an effective two-dimensional contact, related to the three-dimensional one by a geometric factor depending on $\ell_z$. 
\end{abstract}

\maketitle

\paragraph{Introduction.} 

Relating the macroscopic properties of a physical system to microscopic interactions and degrees of freedom is one of the main goals of many-body quantum physics. In ultracold atomic gases not all details of the interaction potential between particles are required since low-energy collisions are generally fully described by the $s$-wave scattering length. As a result, the equation of state of a dilute gas takes a simple, universal, expression where the microscopic physics enters only through two parameters, the mass of the particles and the scattering length. Considering the latter as an additional thermodynamic variable, besides the usual variables (e.g. the chemical potential $\mu$ and the temperature $T$ in the grand canonical ensemble), one can define its thermodynamic conjugate, the so-called Tan two-body contact $C$~\cite{Tan08a,Tan08b,Tan08c}. In a dilute gas, the contact relates the (universal) low-temperature thermodynamics to the (universal) short-distance behavior which shows up in the two-body correlations or the momentum distribution function~\cite{Tan08a,Tan08b,Tan08c,Braaten08,Zhang09,Combescot09,Valiente12,Werner12a,Werner12}. This simple description fails in a strongly interacting Bose gas where other parameters (e.g. associated with three-body effective interactions) are required for a complete description of the universal thermodynamics and short-distance physics~\cite{[{In two dimensions and below, the equation of state can be expressed only in terms of the scattering length and no additional parameters such as the three-body parameter are necessary, see }] Adhikari95}.  

There have been few measurements of the two-body contact in Bose gases. Apart from experiments in the thermal regime~\cite{Wild12,Fletcher17} or the quasi-pure BEC one~\cite{Wild12,Lopes17}, the two-body contact has been determined in a planar Bose gas in a broad temperature range including the normal and superfluid phases as well as the vicinity of the Berezinskii-Kosterlitz-Thouless (BKT) transition~\cite{Zou21}. The experimental data are in good agreement with theoretical predictions in the high-temperature limit (normal gas) and in the low-temperature limit (strongly degenerate superfluid). On the other hand there is no theoretical explanation for the value of the contact obtained near the BKT transition. In particular, the experimental data seem at odds with the predictions of a classical field theory~\cite{Prokofev02} used earlier successfully for the equation of state of a two-dimensional Bose gas~\cite{Yefsah11,Desbuquois14}. 

In this Letter we compute the two-body contact in a  planar Bose gas confined by a harmonic potential using the nonperturbative functional renormalization group (FRG), a modern implementation of Wilson's RG~\cite{Berges02,Delamotte12,Dupuis_review}. This approach has proven to be very accurate for determining the equation of state of such a system~\cite{Rancon12b}. We consider the weak-coupling limit where the three-dimensional scattering length $a_3$ of the bosons is much smaller than the characteristic length $\ell_z=\sqrt{\hbar/m\omega_z}$ of the harmonic potential. We use the three-dimensional thermodynamic definition of the contact where the latter is expressed as a derivative of the pressure with respect to the three-dimensional scattering length $a_3$ and pay special attention to the quasi-two-dimensional structure of the system. We show that the short-distance behavior of the pair distribution function and the high-momentum behavior of the momentum distribution are determined by two contacts: the three-dimensional contact for length scales smaller than $\ell_z$ and, for larger length scales, an effective two-dimensional contact (obtained from the derivative of the pressure with respect to the effective two-dimensional scattering length of the confined bosons), related to the three-dimensional one by a geometric factor depending on $\ell_z$.  We then compare our results with the experimental data of Ref.~\cite{Zou21}. Without any free parameters, we find a remarkable agreement between theory and experiment from low to high temperatures, including the vicinity of the BKT phase transition. We also show how to reconcile these experimental data with the classical field simulations of Ref.~\cite{Prokofev02}.

\paragraph{Contact of a planar Bose gas.} We consider a quasi-two-dimensional system of surface $L^2$  obtained by subjecting a three-dimensional Bose gas to a confining harmonic potential of frequency $\w_z=\hbar/m\ell_z^2$ along the $z$ direction. In the low-temperature regime $k_BT\ll\hbar\w_z$, the physical properties of the planar gas can be obtained from the effective two-dimensional Hamiltonian (from now on we set $\hbar=k_B=1$)
\begin{equation} 
	\hat H = \int d^2r  \biggl\{\frac{\nablabf\hat\psi^\dagger\cdot\nablabf\hat\psi }{2m} 
	+ \frac{g}{2} \hat\psi^\dagger \hat\psi^\dagger \hat\psi \hat\psi  \biggr\},
	\label{ham} 
\end{equation} 
with an ultraviolet momentum cutoff $\Lambda\simeq 0.54\, \ell_z^{-1}$~\cite{Petrov01,Lim08}. The effective interaction constant $g$ is related to the three-dimensional scattering length $a_3$ by
\beq 
mg = \sqrt{8\pi}\frac{a_3}{\ell_z} .
\label{mgdef} 
\eeq 
Computing the low-energy scattering amplitude from~(\ref{ham}), one obtains the effective two-dimensional scattering length $a_2$ as a function of the microscopic parameters of the gas~\cite{Petrov01,Lim08,Pricoupenko07}, 
\begin{equation} 
	\frac{a_2}{\ell_z} \simeq 3.71 \, e^{ -\sqrt{\frac{\pi}{2}}\frac{\ell_z}{a_3}  -\gamma } ,
	\label{a2def}  
\end{equation}
where $\gamma\simeq 0.577$ is the Euler constant. 

In the low-temperature regime $T\ll\w_z$, the pressure can be written in the scaling form characteristic of a two-dimensional system~\cite{Rancon12b}, 
\beq 
\Pquasi(\mu,T) = - \frac{\Oquasi(\mu,T)}{L^2} = \frac{T}{\lambda^2} \calF\left( \frac{\mu}{T}, \tilde g(T)\right) , 
\label{Pdef} 
\eeq 
where $\Oquasi(\mu,T)$ is the grand potential, $\calF$ a universal scaling function, $\lambda=\sqrt{2\pi/mT}$ the thermal de Broglie wavelength and 
\beq 
\tilde g(T) = -\dfrac{4\pi}{\ln\left( \half \sqrt{2ma_2^2 T}\right) + \gamma} 
\eeq  
a temperature-dependent dimensionless interaction constant. Equation~(\ref{Pdef}) is valid for $|\mu|,T\ll \Lambda^2/2m\sim\w_z$. The corrections to the scaling form~(\ref{Pdef}) are negligible when $ma_2^2T,ma_2^2|\mu|\ll 1$. The dependence of the scaling function $\calF$ on $\mu/T$ and $ma_2^2T$ can be simply obtained by dimensional analysis using the fact that the scattering length is the only characteristic length scale at low energies. Renormalization-group arguments show that the dependence on $ma^2_2T$ arises only through $\tilde g(T)$~\cite{Rancon12b}. 

The pressure depending only on $a_3$ (through $a_2\equiv a_2(a_3,\ell_z)$) and not on the details of the interaction potential between particles, it is natural to consider $a_3$ as an additional thermodynamic variable besides $\mu$ and $T$~\cite{Tan08a}. The quasi-two-dimensional two-body contact $\cquasi$ is then essentially defined as the conjugate variable to $a_3$, 
\beq 
\frac{\cquasi}{L^2} =  8\pi m \frac{\partial\Pquasi}{\partial(1/a_3)}\biggl|_{\mu,T} .
\label{Cdef}
\eeq 
$\cquasi$ is an extensive quantity with dimension 1/length. 
It can equivalently be defined in the canonical ensemble by replacing the pressure by the energy density in~(\ref{Cdef}) and taking the derivative at fixed particle number and entropy.  The motivation for defining the contact from a derivative with respect to $1/a_3$ as in a three-dimensional system, rather than with respect to $\ln a_2$ as in a two-dimensional system, is that in the limit $a_3\ll\ell_z$ the collisions keep their three-dimensional character at length scales smaller than $\ell_z$. 

Using the equation of state~(\ref{Pdef}), we can write the contact as 
\beq 
\frac{\cquasi}{L^2} = - 4(2\pi)^{5/2} \frac{T\ell_z}{\lambda^4}\calF^{(0,1)}\left(\frac{\mu}{T},\tilde g(T) \right) \tilde g'(T) ,
\label{Cdef1}
\eeq 
where we use the notation $\calF^{(i,j)}(x,y)=\partial^i_x\partial^j_y \calF(x,y)$. On the other hand, the scaling function  $\calF$ determines the two-dimensional particle and entropy densities, $\bar n=\partial \Pquasi/\partial\mu$ and $s=\partial \Pquasi/\partial T$. This allows us to express $\calF^{(0,1)}$ in terms of $\Pquasi,\bar n,s$, which leads to a relation between the contact and the thermodynamic potentials $\Pquasi$ and $\Eqd$ (as in isotropic systems~\cite{Tan08b})  
\beq 
\frac{\cquasi}{L^2} = 4(2\pi)^{5/2} \frac{\ell_z}{\lambda^2 T} \left( \Pquasi - \frac{\Eqd}{L^2} \right) ,
\label{Cdef2} 
\eeq
where $\Eqd/L^2=-\Pquasi+Ts+\mu\bar n$ is the two-dimensional energy density. 

In the weak-coupling limit, the scattering length $a_2$ is exponentially small with respect to $\ell_z$ and $\tilde g(T)\simeq \tilde g=2mg$ is nearly temperature independent except for exponentially small temperatures $T\sim \w_z e^{-\sqrt{\pi/2}(\ell_z/a_3)}$. For this reason it is often concluded that the two-dimensional Bose gas exhibits an approximate scale invariance~\cite{Prokofev02,Yefsah11,Desbuquois14} (with no characteristic energy scales other than $\mu$ and $T$): $\calF(\mu/T,\tilde g(T))\simeq \calF(\mu/T,\tilde g)$~\cite{Rancon12b}, and the normalized contact $\cquasi\lambda^4/L^2\ell_z$ is a function of $\mu/T$ and $\tilde g$. 
However the assumption of approximate scale invariance cannot be used in the calculation leading to~(\ref{Cdef2}) since this would imply $\tilde g'(T)=0$ and in turn $\Eqd= L^2 \Pquasi$ and $\cquasi=0$. The contact defined in~(\ref{Cdef2}) can thus be seen as a measure of the breakdown of scale invariance in the planar Bose gas~\cite{not55}. The situation is different in an isotropic three-dimensional system; in the unitary limit $a_3\to \infty$ where scale invariance is satisfied, $E_{\rm 3D}=(3/2)L^3P_{\rm 3D}$, but the contact remains finite since $(3/2)P_{\rm 3D}-E_{\rm 3D}/L^3=\calO(1/a_3)$ whereas $\cthree\propto a_3[(3/2)P_{\rm 3D}-E_{\rm 3D}/L^3]$. 

A remarkable feature of the contact defined from the pressure is that it also determines the short-distance behavior of the pair distribution function as well as the high-momentum limit of the momentum distribution function~\cite{Tan08a,Tan08b,Tan08c,Braaten08,Zhang09,Combescot09,Valiente12,Werner12a,Werner12}. This property still holds in the planar gas. For the pair distribution function, averaged over the position $\R=(\r_1+\r_2)/2$ of the pair center of mass, one has~\cite{notSM} 
\begin{align} 
	g(\r) &= \int d^3R\, \mean{\hat\psi^\dagger(\r_1)\hat\psi^\dagger(\r_2) \hat\psi(\r_2) \hat\psi(\r_1)}, \nonumber \\ 
	&= \llbrace 
	\begin{array}{lll} \frac{\cquasi}{(4\pi)^2}\left(\frac{1}{|\r|}-\frac{1}{a_3}\right)^2 & \mbox{if} & |\r|\ll \ell_z, \lambda, d , \\ 
		\frac{\cquasi}{\sqrt{2\pi}\ell_z} |\phi_0^-(z)|^2 \left( \frac{\ln|\rhobf/a_2|}{2\pi} \right)^2 & \mbox{if} & \ell_z\ll |\rhobf|\ll \lambda, d ,  
	\end{array}
	\right. 
\end{align}
where $\r=\r_1-\r_2=(\rhobf,z)$ is the coordinate of the relative motion of the pair (with $\rhobf$ a two-dimensional coordinate) and $d$ the mean interparticle distance. $\phi_0^-$ denotes the ground state of a particle of reduced mass $m/2$ in an harmonic potential of characteristic frequency $\w_z$. The fact that the contact~(\ref{Cdef}) determines the pair distribution function at short length scales, $|\r|\ll \lambda,\ell_z,d$, is not surprising since $\cquasi\propto \partial \Pquasi/\partial(1/a_3)$ is defined as in the three dimensional case. On the other hand, the effective contact that appears in the regime $\ell_z\ll |\rhobf|\ll \lambda,d$ can be written as
\beq 
\frac{\cquasi}{\sqrt{2\pi}\ell_z L^2} = -4\pi m \frac{\partial \Pquasi}{\partial \ln a_2} ,
\eeq 
and thus coincides with the usual thermodynamic definition of a two-dimensional contact. Similarly, for the momentum distribution function $\bar n_{\k,k_z}$, one finds
\beq
\bar n_{\k,k_z} =  \frac{\cquasi}{(\k^2+k_z^2)^2} \quad \mbox{if} \quad \frac{1}{a_3},\frac{1}{d},\frac{1}{\lambda}  \ll \sqrt{\k^2+k_z^2}, 
\label{Cnka} 
\eeq 
and 
\begin{align}
	\bar n^{\rm (2D)}_{\k} &= \int\frac{dk_z}{2\pi} \bar n_{\k,k_z} ,\nonumber \\ 
	&= \frac{1}{\sqrt{2\pi}\ell_z} \frac{\cquasi}{|\k|^4} \quad \mbox{if} \quad \frac{1}{d}, \frac{1}{\lambda}  \ll |\k| \ll  \frac{1}{\ell_z}, 
	\label{Cnkb} 
\end{align}
where we assume the normalization $\frac{1}{L^2}\sum_{\k}\int\frac{dk_z}{2\pi}\bar n_{\k,k_z}=N$ (with $N\equiv\mean{\hat N}$ the total number of particles). Equation~(\ref{Cnka}) holds at any temperature while Eq.~(\ref{Cnkb}) is valid only in the low-temperature regime $T\ll\w_z$. Note that a result similar to~(\ref{Cnka},\ref{Cnkb}) has been obtained for fermions  in quasi-one- and quasi-two-dimensional traps~\cite{He19} (see also \cite{Decamp2018,Bougas20} for bosons in a quasi-one-dimensional trap).

\paragraph{FRG calculation of the contact.} 

Following Ref.~\cite{Rancon12b} we compute the pressure of the planar Bose gas at low temperatures $T\ll\w_z$ using the two-dimensional Hamiltonian~(\ref{ham}). The main quantity of interest in the FRG approach is the effective action (or Gibbs free energy) $\Gamma$, defined as the Legendre transform of the Helmholtz free energy, and which is directly related to the pressure: $P=-(T/L^2)\Gamma$. We refer to Refs.~\cite{Rancon12b,Dupuis_review} for a detailed description of the method. Once the pressure is known, the contact is obtained from~(\ref{Cdef}).  

\begin{figure}
	\centerline{\includegraphics[width=8.75cm]{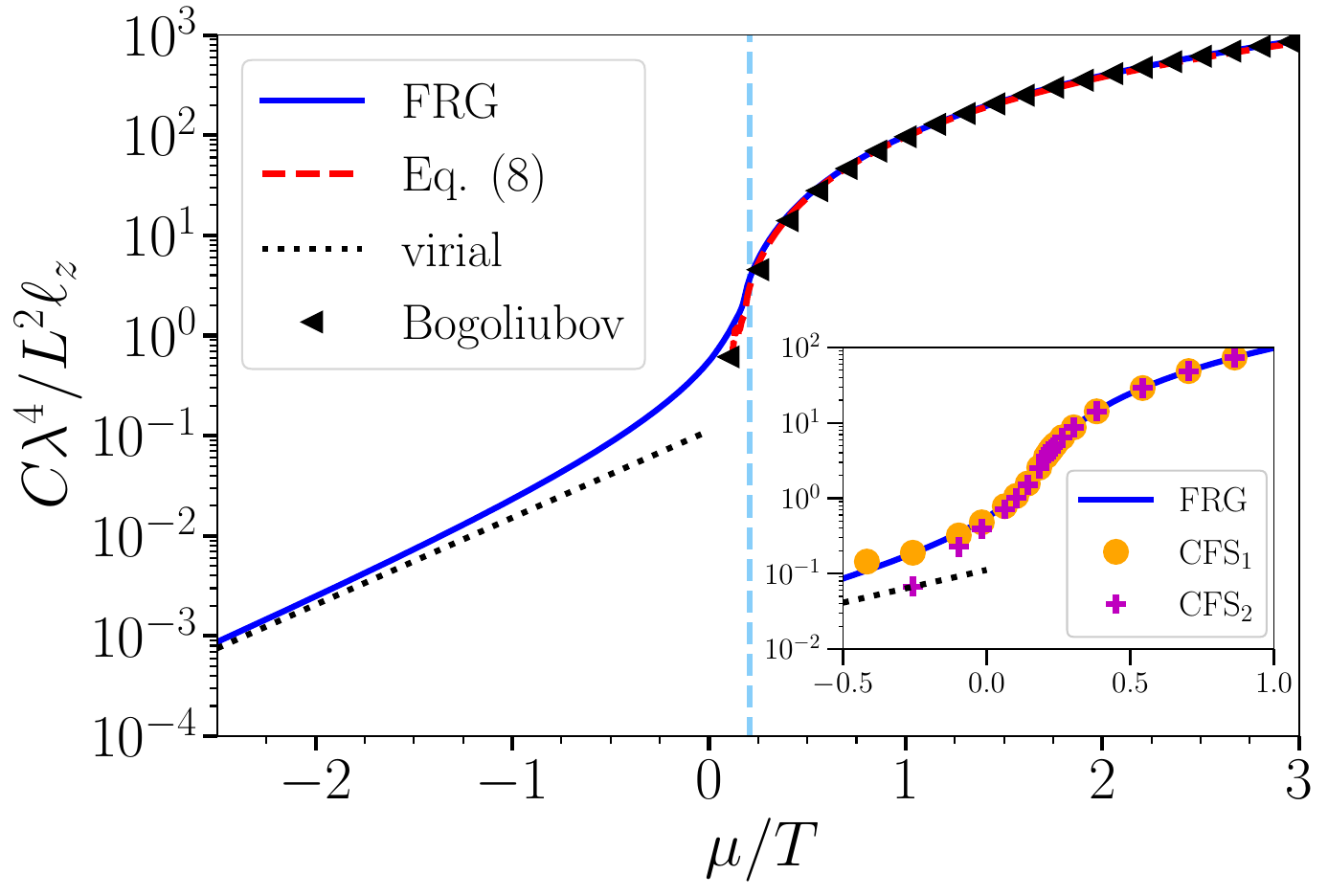}}
	\caption{Two-body contact $\cquasi\lamb^4/L^2\ell_z$ {\it vs} $\mu/T$ for $2mg=0.32$ and $T/\w_z\simeq 0.00145$ as obtained from the FRG (solid (blue) line). The dashed (red) line shows the result obtained from~(\ref{Cdef2}) while the symbols correspond to the Bogoliubov result~(\ref{Cbog}) and the virial expansion~(\ref{Cvirial}). The BKT transition occurs  at $\mu/T\simeq 0.21$, shown as a vertical dashed line, as estimated from FRG. The inset shows a comparison between FRG and classical field simulations using the thermodynamic definition~(\ref{Cdef}) (CFS${}_1$, (orange) circles) and the relation between $C$ and $\mean{(\hat\psi^\dagger)^2(\hat\psi)^2}$ (CFS${}_2$, (purple) crosses); see text.}
	\label{fig_Cth}
\end{figure}

The FRG approach to interacting boson systems has proven to be very accurate. In particular, 
the universal scaling function $\calF$ entering the equation of state~(\ref{Pdef}) of the two-dimensional Bose gas has been computed using the FRG approach~\cite{Rancon12b}, and very good agreement with experimental data in planar gases~\cite{Hung11,Yefsah11,Zhang12}, with or without an optical lattice, has been obtained~\cite{not20}. 

In Fig.~\ref{fig_Cth}, we show the contact $\cquasi$ normalized by $L^2\ell_z/\lambda^4$ obtained for $2mg =2\sqrt{8\pi}a_3/\ell_z=0.32$ and $T/\w_z\simeq 0.00145$ by computing the pressure $\Pquasi$ for two nearby values of $a_3$ and taking a numerical derivative. We find a very good agreement with two limiting cases~\cite{notSM}, the zero-temperature limit in the superfluid phase where 
\beq 
\frac{C_{\rm Bog}}{L^2} = \sqrt{2\pi} (m\mu)^2 \ell_z 
\label{Cbog} 
\eeq 
can be obtained from the Bogoliubov theory (including the Lee-Huang-Yang correction~\cite{not33}), and the dilute  normal gas where 
\beq
\frac{C_{\rm virial}}{L^2} = \frac{z^2\ell_z}{\lambda^4} \frac{2(2\pi)^{5/2}}{\left| \ln\left( \frac{a_2}{\lambda} \sqrt{\frac{\pi}{2}} e^\gamma\right) \right|^2}
\label{Cvirial} 
\eeq
(with $z=e^{\beta\mu}$ the fugacity) can be obtained from the virial expansion. The contact deduced from~(\ref{Cdef2}), the two-dimensional energy density $\Eqd/L^2=-\Pquasi+T \partial \Pquasi/\partial T+\mu \partial \Pquasi/\partial\mu$ being obtained from numerical derivatives, is also shown in the figure. The apparent disagreement for $\mu/T\lesssim 0.5$ is due to a lack of precision in the numerical calculation, the values of $\Pquasi$ and $\Eqd/L^2$ being extremely close and the normalized contact very small. 

The inset of Fig.~\ref{fig_Cth} shows the contact obtained from the classical field simulations of Ref.~\cite{Prokofev02} using two different methods. 
The first one (CFS${}_1$) is based on the thermodynamic definition~(\ref{Cdef}) of the contact and the calculation of the pressure $P$, the second one (CFS${}_2$) uses the definition $\cquasi/L^2=4(2\pi)^{3/2}(a_3^2/\ell_z)\mean{(\hat\psi^\dagger)^2(\hat\psi)^2}$ (which follows from~(\ref{Cdef})). 
Both $P$ and $\mean{(\hat\psi^\dagger)^2(\hat\psi)^2}$ are obtained from the results of Ref.~\cite{Prokofev02} (see~\cite{notSM}). The first method has one fitting parameter, the value of the contact at the BKT transition, which we determine by minimizing the relative difference between the FRG and the classical field results. The first method CFS${}_1$ and the FRG are in agreement with an accuracy better than 1\% when $\mu/T\gtrsim 0.25$. In the range $0\lesssim \mu/T\lesssim 0.25$, which includes the fluctuation region about the BKT transition, the agreement remains within 5\% but deteriorates when $\mu/T<0$. The second method, CFS${}_2$, is clearly much less accurate when $\mu/T<0$ and breaks down in the low-density limit since $C$ becomes negative when $\mu/T\lesssim -0.3$.

\paragraph{Comparison with the experiment.} 

\begin{figure}[t]
	\centerline{\includegraphics[width=8.75cm]{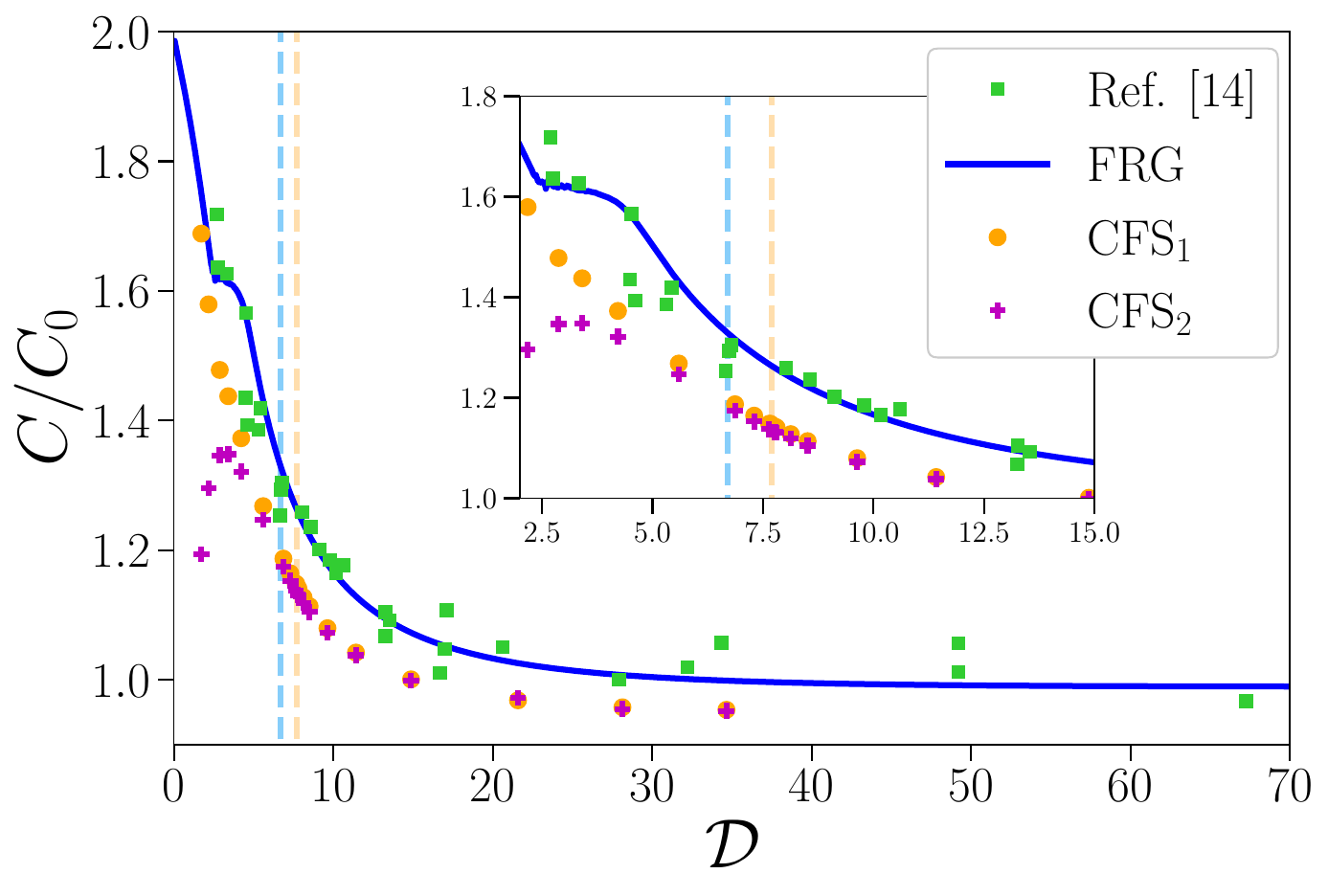}}
	\caption{Normalized two-body contact $\cquasi/C_{0}$ as a function of the phase-space density $\calD=\bar n \lambda^2$ for $2mg=0.32$. The square (green) symbols show the experimental data from Ref.~\cite{Zou21} and the solid (blue) line the FRG result at $T/\w_z=0.03$. The (orange) circles and (purple) crosses are the estimate from classical field simulations, CFS${}_1$ and CFS${}_2$, respectively (see text). The vertical dashed lines correspond to the position of the BKT transition as estimated from FRG at $\calD\simeq 6.7$ (blue line on the left) and classical field simulations at $\calD\simeq 7.7$ (orange line on the right). The inset is a zoom on the fluctuation region about the BKT transition.
	}
	\label{fig_Cexp}
\end{figure}

The thermodynamic definition~(\ref{Cdef}) of the contact was realized experimentally by means of a Ramsey interferometric method~\cite{Zou21}. The measurements were performed on a Bose gas of $^{87}$Rb atoms confined by a harmonic potential with $2mg=2\sqrt{8\pi}a_3/\ell_z=0.32$ in a broad temperature range around the BKT transition. In Fig.~\ref{fig_Cexp} we compare the experimental data with the FRG and classical field results obtained for $2mg=0.32$. Following Ref.~\cite{Zou21} we show the contact normalized by the mean-field contact $C_0=4(2\pi)^{3/2}L^2\bar n^2a_3^2/\ell_z$ as a function of the phase-space density $\calD=\bar n\lambda^2$. The FRG calculation is done at $T/\omega_z=0.03$, low enough to be in the quasi-2D regime ($T\ll \Lambda^2/2m$) but high enough to minimize the logarithmic corrections (in the experiment, $T/\omega_z \in[0.05,0.75]$). The normalized contact varies between~\cite{not34} 
\beq
\frac{C_{\rm Bog}}{C_0} = \frac{\pi}{2} \frac{ \ell_z^2/a_3^2}{\left| \ln \left( a_2 \sqrt{\pi \bar n} e^{\gamma+1/2} \right)\right|^2 } \simeq 1 
\eeq 
for $\calD\gg 1$ 
and 
\beq
\frac{C_{\rm virial}}{C_0} =  \frac{\ell_z^2}{a_3^2}  \frac{\pi}{\left| \ln \left(\frac{a_2}{\lambda} \sqrt{\frac{\pi}{2}} e^\gamma \right) \right|^2} \simeq 2 
\eeq
for $\calD\ll 1$. 

The agreement between FRG and the experimental data is very good in spite of the absence of any fit parameter.  On the other hand, there is a downward shift of the classical field result with respect to the experimental data (only CFS${}_2$ was compared to the experimental data in Ref.~\cite{Zou21}). This does not come from the value of the contact, which is highly accurate for $\mu>0$ as previously discussed (Fig.~\ref{fig_Cth}), but apparently from a lower accuracy of the density estimate and therefore the normalized contact $C/C_0$. 
The relative difference between the FRG and CFS results for the density(for a given value of the chemical potential) is only about 5\%, but this is enough to have a visible effect in the plot of $C/C_0$. Normalizing the contact with $C_{\rm Bog}$ gives the same qualitative results~\cite{notSM}. The classical field result CFS${}_1$, apart from this downward shift, provides us with a good fit of the experimental data except in the low-density limit $\calD\lesssim 1.5$ where it becomes much too large (and therefore does not appear in Fig.~\ref{fig_Cexp}). Note that there is no sign of the BKT transition, predicted to occur around $\calD\simeq 7.7$ (classical field simulations~\cite{Prokofev02}) and $\calD\simeq 6.7$ (FRG), marked as vertical dashed lines  in Fig.~\ref{fig_Cexp}. 
The bump predicted by the FRG near $\calD=3$ is due to the rapid change of $\bar n$, and therefore $C_0$, for small and positive values of $\mu/T$~\cite{Yefsah11,Desbuquois14}. Although this bump also appears in the experimental data, the 3--10\% experimental uncertainty~\cite{not21} does not allow us to assert that it is a real feature.

\paragraph{Conclusion.}

The theoretical calculation of the two-body contact of a planar Bose gas in the framework of the nonperturbative FRG is in remarkable agreement with the experimental data of Ref.~\cite{Zou21}. The FRG provides us with an accurate value of the contact over a broad temperature range including the vicinity of the BKT transition. We have also shown how the apparent contradiction between the experimental data of Ref.~\cite{Zou21} and the classical field simulations~\cite{Prokofev02} can be overcome. The FRG predicts a bump in the contact for a phase-space density $\calD\simeq 3$, which requires improved-precision measurements to be confirmed. 

An interesting aspect of the contact theory in a quasi-two-dimensional gas is that the high-momentum tail, $\sqrt{\k^2+k_z^2}\gg 1/a_3,1/\lambda,1/d$, of the three-dimensional momentum distribution is controlled by the three-dimensional contact (as defined in~(\ref{Cdef})) while the intermediate momentum range, $1/d,1/\lambda\ll |\k|\ll 1/\ell_z$, is controlled by the effective two-dimensional contact $\cquasi/\sqrt{2\pi}\ell_z$ [Eqs.~(\ref{Cnka},\ref{Cnkb}) and Refs.~\cite{He19,Bougas20}]. A measure of the momentum distribution, for instance by  ballistic expansion of the atomic cloud or rf spectroscopy~\cite{Stewart10}, would allow us to confirm this essential property of the contact theory.

\paragraph{Acknowledgment.} We thank J. Beugnon, J. Dalibard, S. Nascimbene for providing us with the experimental data. We are grateful to them as well as  F. Werner for enlightening discussions and a critical reading of the manuscript. AR is  supported  by  the  Research  Grant  QRITiC  I-SITEULNE/ANR-16-IDEX-0004 ULNE.

%



\clearpage

\onecolumngrid

\begin{center}
	\textbf{\large  Tan's two-body contact in a planar Bose gas \\ 
		[.3cm] -- Supplemental Material --} \\ 
	[.4cm] Adam Rançon$^{1,2}$ and Nicolas Dupuis$^3$ \\[.1cm]
	{\itshape $^1$Univ. Lille, CNRS, UMR 8523 – PhLAM – Laboratoire de Physique des Lasers Atomes et Molécules, F-59000 Lille, France} \\ 
{\itshape $^2$Institute  of  Physics,  Bijeni\v cka  cesta  46,  HR-10001  Zagreb,  Croatia}\\
	{\itshape $^3$Sorbonne Universit\'e, CNRS, Laboratoire de Physique Th\'eorique de la Mati\`ere Condens\'ee, LPTMC, F-75005 Paris, France \\}
\end{center}

\setcounter{equation}{0}
\setcounter{figure}{0}
\setcounter{table}{0}
\setcounter{page}{1}
\renewcommand{\theequation}{S\arabic{equation}}
\renewcommand{\thefigure}{S\arabic{figure}}	
\renewcommand{\bibnumfmt}[1]{[S#1]}
\renewcommand{\citenumfont}[1]{S#1}

In the Supplemental Material, we discuss the calculation of the two-body contact of a quasi-two-dimensional (quasi-2D) dilute Bose gas in the zero-temperature superfluid phase within the Bogoliubov theory and in the normal (non-degenerate) phase within the virial expansion (Secs.~I and II). In particular the virial expansion shows that the short-distance behavior of the pair distribution function and the high-momentum behavior of the momentum distribution are determined by two contacts: the 3D contact for length scales smaller than $\ell_z$ and, for larger length scales, an effective 2D contact (obtained from the derivative of the pressure with respect to the effective 2D scattering length of the confined bosons), related to the 3D one by a geometric factor depending on $\ell_z$. We generalize these results beyond the virial expansion in Sec.~II.D. Finally we show how the theoretical value of the contact can be deduced from the classical field simulations of Ref.~\cite{Prokofev02_sm} (Sec.~III).  

We start from the (grand canonical) Hamiltonian 
\beq  
\hat H = \int d^3r  \biggl\{\frac{\nablabf\hat\psi^\dagger(\r)\cdot\nablabf\hat\psi(\r) }{2m} + [V(z)- \mu] \hat\psi^\dagger(\r) \hat\psi(\r) 
+ \frac{g_3}{2} \hat\psi^\dagger(\r) \hat\psi^\dagger(\r) \hat\psi(\r) \hat\psi (\r) \biggr\} ,
\eeq 
where $V(z)=\half m\w_z^2 z^2$ is a harmonic trapping potential along the $z$ direction and $\ell_z=1/\sqrt{m\w_z}$ is the width of the trap (we set $\hbar=k_B=1$). The surface $L^2$ of the quasi-2D system is assumed much larger than all characteristic length scales of the system. We denote by $g_3$ the interaction between bosons, assumed to be local, and regularize the model by a UV momentum cutoff $\Lambda_3$. In the absence of the trapping potential ($V(z)=0$), the $s$-wave scattering length is defined by 
\beq 
\frac{m}{4\pi a_3} = \frac{1}{g_3} +\int_{\k,k_z} \frac{1}{2 \eps_{{\bf k},k_z}} ,
\label{a3def} 
\eeq 
where  $\eps_{{\bf k},k_z}=(\k^2+k_z^2)/2m$, $\int_{\k,k_z}=\int \frac{d^2k}{(2\pi)^2}\int \frac{dk_z}{2\pi}$ (with $\k$ a 2D wavevector) and the momentum integral is restricted to $(\k^2+k_z^2)^{1/2}\leq\Lambda_3$.

In presence of the trapping potential, scattering at low energy is effectively 2D (see Sec.~II.A) with the 2D $s$-wave scattering length~\cite{Petrov01_sm,Lim08_sm,Pricoupenko07_sm} 
\beq
a_2 \simeq 3.71 \ell_ze ^{-\sqrt{\frac\pi2}\frac{\ell_z}{a_3}-\gamma},
\label{app0}
\eeq
where $\gamma\simeq 0.577$ is the Euler constant. 
In the low-temperature regime $T\ll 1/m\ell_z^2$, where $\ell_z$ is much smaller than the thermal de Broglie wavelength $\lambda=\sqrt{2\pi/mT}$, only the lowest-energy state of the trapping potential $V(z)$ is occupied and the low-energy properties of the quasi-2D gas can be obtained from the effective 2D Hamiltonian
\begin{equation} 
	\hat H_{\rm eff} = \int d^2\rho  \biggl\{\frac{\nablabf\hat\psi^\dagger(\rhobf)\cdot\nablabf\hat\psi(\rhobf) }{2m} 
	- \mu \hat\psi^\dagger(\rhobf) \hat\psi(\rhobf)  + \frac{g}{2} \hat\psi^\dagger(\rhobf) \hat\psi^\dagger(\rhobf) \hat\psi(\rhobf) \hat\psi(\rhobf)  \biggr\}
	\label{app1} 
\end{equation} 
with an ultraviolet momentum cutoff $\Lambda$. Computing the low-energy scattering amplitude from~(\ref{app1}), one obtains the effective 2D scattering length $a_2$ as a function of the effective parameters $g$ and $\Lambda$  
\begin{equation} 
	a_2  = \frac{2}{\Lambda} e^{ -\frac{2\pi}{mg}  -\gamma }.
	\label{app3}  
\end{equation}
This reproduces the scattering length~\eqref{app0} setting $\Lambda\simeq 0.54\, \ell_z^{-1}$ and
\beq 
mg = \sqrt{8\pi}\frac{a_3}{\ell_z} . 
\label{g2D} 
\eeq 

In the planar Bose gas, we define the contact as in a 3D system, 
\beq 
\frac{C}{L^2} = 8\pi m \frac{\partial P}{\partial (1/a_3)} ,
\label{app24} 
\eeq 
i.e. from the derivative of the pressure with respect to the scattering length $a_3$.

\section{I.\;\; Zero temperature: Bogoliubov theory} 

In the zero-temperature degenerate limit ($\mu>0$) we expect most particles to be in the condensate. In that case the contact should be well described by the effective 2D Hamiltonian~(\ref{app1}) and the Bogoliubov theory. 

Following Ref.~\cite{Rancon12d_sm} we consider the effective potential (or Gibbs free energy) $U(n)$ where $n$ is the 2D condensate density. At the mean-field level, it is simply given by the Hamiltonian treating the operator $\hat\psi\equiv \sqrt{n}e^{i\theta}$ (with $\theta$ an arbitrary phase) as a classical field, 
\beq 
U(n) = - \mu n + \frac{g}{2} n^2 ,
\eeq  
assuming $n$ to be uniform. The equilibrium value of the condensate density, $n_0=\mu/g$, is obtained by minimizing the effective potential: $U'(n_0)=0$. This gives the pressure $P=-U(n_0)=\mu^2/2g$ and the particle density $\bar n=\partial P/\partial\mu=\mu/g$ at mean-field. We thus obtain the contact 
\begin{equation}
	\frac{C}{L^2} = \sqrt{2\pi} (m\mu)^2 \ell_z . 
	\label{bog0}
\end{equation}

The Lee-Huang-Yang correction to the mean-field result is given by the one-loop expression~\cite{Rancon12d_sm}
\beq 
U(n) = - \mu n + \frac{g}{2} n^2 + \frac{1}{2\ell} \int_\k [  -\xik  - 2gn + \Ek(n) ] , 
\label{bog1}
\eeq 
where $\xik=\epsk-\mu$  and 
\beq 
\Ek(n) = \bigl[ (\xik+2gn)^2 - (gn)^2 \bigr]^{1/2} .
\eeq 
The parameter $\ell$ in~(\ref{bog1}) is necessary to organize the loop expansion but will eventually be set to unity. From~(\ref{bog1}) we obtain
\beq 
\begin{split} 
	U'(n) &= - \mu + gn - \frac{g}{\ell} \int_\k \left( 1 - \frac{\Ek'(n)}{2g} \right) , \\ 
	n_0 &= \frac{\mu}{g} + \frac{1}{\ell} \int_\k \left( 1 - \frac{2\epsk+\mu}{2\Ek} \right) ,
\end{split}
\eeq 
where it is now understood that all calculations are performed to order $1/\ell$ and $\Ek=\Ek(n=\mu/g)=[\epsk(\epsk+2\mu)]^{1/2}$ is the value of $\Ek(n_0)$ to leading order. Performing the momentum integration with the UV cutoff $\Lambda$, we obtain 
\beq
\begin{split} 
	n_0 &= \frac{\mu}{g} + \frac{1}{\ell}  \left[ \frac{m\mu}{4\pi} \ln\left( \frac{\Lambda^2}{m\mu} \right) - \frac{m\mu}{2\pi} \right] ,  \\
	P &= \frac{\mu^2}{2g} - \frac{1}{2\ell} \int_\k (\Ek-\epsk-\mu) 
	= \frac{\mu^2}{2g} - \frac{1}{\ell} \frac{m}{4\pi} \left[ \frac{\mu^2}{4} - \frac{\mu^2}{2} \ln \left( \frac{\Lambda^2}{m\mu} \right) \right] . 
\end{split}
\eeq 
Using the definition~(\ref{app3}) of the scattering length $a_2$ and setting $\ell=1$, we deduce~\cite{Popov_book2_sm,Mora03_sm,Pricoupenko04_sm}
\beq
\begin{split} 
	P &= - \frac{m\mu^2}{8\pi} \ln \left( \frac{m\mu a_2^2}{4} e^{2\gamma+\half} \right) , \\
	\bar n &= - \frac{m\mu}{4\pi} \ln \left( \frac{m\mu a_2^2}{4} e^{2\gamma+1} \right) .
\end{split}
\label{bog2} 
\eeq  
The contact as a function of $\mu$ retains the same expression as {in the mean-field approximation [Eq.~(\ref{bog0})], but the relation between $\mu$ and the density $\bar n$ differs from the mean-field one. From~(\ref{bog2}) and (\ref{bog0}) we obtain
	\beq 
	\mu \simeq \frac{2\pi \bar n/m}{\left| \ln\left(a_2 \sqrt{\pi \bar n}e^{\gamma+\half}\right) \right|},
	\eeq 
	and the zero-temperature value of the contact,
	\beq 
	\frac{C_{\rm Bog}}{L^2} \simeq (2\pi)^{5/2}  \frac{\bar n^2 \ell_z}{\left| \ln\left(a_2 \sqrt{\pi \bar n}e^{\gamma+\half}\right) \right|^2} ,
	\label{CBog}
	\eeq 
	as a function of the density.
	
	From~(\ref{bog2}) it is straightforward to recover the expression $C/L^2=\sqrt{2\pi}(m\mu)^2\ell_z$ using Eq.~(8) in the main text relating the contact to the pressure and the energy density. This is not the case in the mean-field theory; the latter gives a vanishing contact using~(8) but a nonzero one using~(\ref{app24}).

	\section{II.\;\; Low-density limit: virial expansion} 
	
	When the gas is in the normal (non-degenerate) phase, $\mu<0$ and $|\mu|\gg T$, it is possible to compute the contact within an expansion in the fugacity $z=e^{\beta\mu}$. This calculation not only gives an explicit expression of the contact in that limit but also shows that the short-distance behavior of the pair distribution function and the high-momentum behavior of the momentum distribution are determined by two contacts: the  3D contact for length scales smaller than $\ell_z$ and an effective 2D contact for length scales larger than $\ell_z$. We show that the relation between these two contacts is valid beyond the virial expansion.

	\subsection{A. Two-body problem and $T$ matrix \label{sec_2body}} 
	
	The Hamiltonian of a single particle in the harmonic potential reads 
	\beq 
	\hat H = \frac{\hat\p^2}{2m} +  \frac{\hat p_z^2}{2m} +\half m \w_z^2 \hat z^2 ,
	\eeq 
	where $\p$ is a 2D momentum associated with the motion perpendicular to the $z$ axis and we denote the 3D coordinate by $\r=(\rhobf,z)$, while $\k$ is a 2D wavevector. The eigenstates $\ket{\k,n}$ and eigenenergies are given by 
	\beq 
	\braket{\r}{\k,n} = \frac{e^{i\k\cdot\rhobf}}{L} \phi_{n}(z), \qquad 
	\eps_{\k,n} = \eps_\k + \eps_n = \frac{\k^2}{2m} +  \left( n+ \half \right) \w_z 
	\eeq 
	with 
	\beq 
	\phi_{n}(z) = 
	\frac{1}{\sqrt{2^n n!}} \left( \frac{m\w_z}{\pi}\right)^{1/4} H_n \left(\sqrt{m\w_z} z \right) e^{-\frac{m\w_z}{2} z^2 } 
	\eeq 
	where $H_n$ is a Hermite polynomial ($n\in\mathbb N$). For the two-body problem it is convenient to introduce the center-of-mass variables and those of the relative motion,  
	\beq
	\begin{split}
		&\hat\P = \hat\p_1 + \hat\p_2 ,  \quad \hat P_z = \hat p_{z1} + \hat p_{z2} , \quad 
		\R_\para = \frac{\rhobf_1+\rhobf_2}{2} , \quad 
		\hat Z = \frac{\hat z_1 + \hat z_2}{2} , \\ 
		&\hat\p = \frac{\hat\p_1 - \hat\p_2}{2} , \quad \hat p_z = \frac{\hat p_{z1} - \hat p_{z2}}{2} , \quad  \rhobf = \rhobf_1 - \rhobf_2 , \quad 
		\hat z=\hat z_1 - \hat z_2  ,
	\end{split} 
	\eeq 
	and write the two-body  Hamiltonian as 
	\begin{align}
		\hat H ={}& \frac{\hat\P^2}{4m} + \frac{\hat P_z^2}{4m} + \half (2m) \w_z^2 \hat Z^2 
		+ \frac{\hat\p^2}{m} + \frac{\hat p_z^2}{m} +  \half \left(\frac{m}{2}\right) \w_z^2 \hat z^2  + g_3 \,  \delta(\hat\rhobf) \delta(\hat z) .
	\end{align}
	The harmonic potential preserves the decoupling between the center-of-mass and relative motions. The two-body eigenstates for $g_3=0$ can therefore be written as $\ket{\k_1,n_1;\k_2,n_2}=\ket{\k_1,n_1}\otimes\ket{\k_2,n_2}$ with the corresponding (non-symmetrized) wavefunctions  
	\beq 
	\frac{e^{i\k_1\cdot\rhobf_1 + i\k_2\cdot \rhobf_2}}{L^2} \phi_{n_1}(z_1)  \phi_{n_2}(z_2) , \qquad \eps_{\k_1,n_1,\k_2,n_2} = \eps_{\k_1,n_1} + \eps_{\k_2,n_2} \qquad (n_1,n_2 \in \mathbb{N}) 
	\eeq 
	or, in the center-of-mass frame, $\ket{\K,N;\k,n}$ with the corresponding wavefunctions
	\beq 
	\frac{e^{i\K\cdot\R_\para + i\k\cdot \rhobf}}{L^2} \phi^+_{N}(Z)  \phi^-_{n}(z) , \qquad \eps_{\K,N,\k,n} = \eps^+_{\K,N} + \eps^-_{\k,n} \qquad (N,n \in \mathbb{N}) ,
	\eeq 
	where $\phi_n^\pm,\eps^\pm_{\k,n}$ are deduced from $\phi_n,\eps_{\k,n}$ by replacing the mass $m$ by $2m$ and $m/2$, respectively. 
	
	In the center-of-mass frame, the $T$ matrix reads 
	\beq 
	T_{nn'}(\K,N,\tau-\tau') = \bra{\K,N;\k,n} \hat T\ket{\K,N;\k',n'} = \Theta(\tau-\tau') e^{-\eps^+_{\K,N}(\tau-\tau')} t_{nn'}(\tau-\tau') 
	\eeq 
	i.e. 
	\beq 
	T_{nn'}(\K,N,i\w) = t_{nn'}(i\w-\eps^+_{\K,N}) , 
	\eeq 
	where $\w$ is a Matsubara frequency (we suppress the index $n$ of $\wn$ since we consider here the zero-temperature limit). The part of the $T$ matrix that describes the relative motion can be easily obtained from the Lippmann-Schwinger equation~\cite{Lim08_sm}, 
	\beq 
	t_{nn'}(i\w) = \phi_n^-(0)^*  \phi_{n'}^-{}(0) t(i\w) , \qquad 
	\frac{1}{t(i\w)} = \frac{1}{g_3} - \sum_{n} \int_\k \frac{|\phi_n^-(0)|^2}{i\w-\eps^-_{\k,n}} . 
	\label{app10} 
	\eeq  
	Here and in the following the sum over $n$ extends over $\mathbb{N}$. 
	We can use the definition~(\ref{a3def}) of the scattering length $a_3$ of the unconfined bosons
	to rewrite~(\ref{app10}) as
	\beq \frac{1}{t(i\w)} = \frac{m}{4\pi a_3}  - \sum_n \int_\k \frac{|\phi_n^-(0)|^2}{i\w-\eps^-_{\k,n}} - \int_{\k,k_z} \frac{1}{2(\eps_\k + \eps_{k_z})} ,
	\label{app11}  
	\eeq  
	where the UV cutoff $\Lambda_3$ can now be sent to infinity~\cite{Lim08_sm}. 
	At low frequencies, $|\w|\ll \w_z$, the $T$ matrix $t_{00}(i\w)$ takes the usual 2D form~\cite{Petrov00a_sm,Lim08_sm}
	\beq 
	t_{00}(i\w) = - \frac{2\pi/m}{\ln \frac{a_2}{2}\sqrt{-im\w}+\gamma} ,  
	\label{app23}  
	\eeq 
	where the effective 2D scattering length $a_2$ is given by~(\ref{app3}). Finally we note that Eq.~(\ref{app11}) implies that the retarded $T$ matrix $t^R(\w)=t(\w+i0^+)$ satisfies 
	\begin{align}
		&\frac{\partial }{\partial \w} \frac{1}{t^R(\w)} =  \sum_n \int_\k \frac{|\phi_n^-(0)|^2}{(\w+i0^+ -\eps^-_{\k,n})^2} ,
		\label{app12} \\
		&\frac{\partial }{\partial (1/a_3)} \frac{1}{t^R(\w)} = \frac{m}{4\pi} . \label{app13}
	\end{align}  
	These two relations will be useful in the calculation of the contact.

	\subsection{B. Calculation of the contact from the thermodynamic definition} 
	
	Let us consider the distribution of the in-plane momentum $\bar n_\k$. To order $z^2$, and ignoring the part that comes from the non-interacting propagator $G(\k,n,i\wnu)=(i\wnu+\mu-\eps_{\k,n})^{-1}$ and does not depend on the interactions, it is sufficient to consider~\cite{Leyronas11_sm} 
	\beq 
	\delta \bar n_\k = \frac{2}{\beta^2} \sum_{\wn,\wnu} \sum_{n_1,n_2}\sum_{\K} G(\k,n_1,i\wn)^2 G(\K-\k,n_2,i\wnu-i\wn) T_{n_1,n_2;n_1,n_2}(\K,i\wnu) , 
	\eeq 
	where 
	\begin{align}
		T_{n_1,n_2;n_1,n_2}(\K,i\wnu) &= \bra{\k_1,n_1;\K-\k_1,n_2} \hat T\ket{\k'_1,n_1;\K-\k'_1,n_2} \nonumber\\ 
		&= \sum_{N,n} U_{n_1,n_2;N,n} T_{nn}(\K,N,i\wnu) U^\dagger_{N,n;n_1,n_2} \nonumber\\ 
		&= \sum_{N,n} |U_{n_1,n_2;N,n}|^2 |\phi_n^-(0)|^2 t(i\wnu-\eps^+_{\k,N}+2\mu) 
	\end{align}  
	is obtained using the eigenstates $\ket{\k_1,n_1;\k_2,n_2}=\ket{\k_1,n_1}\otimes\ket{\k_2,n_2}$ for the two-body transverse motion. We have used the fact that $U_{n_1,n_2;N,n}=\braket{n_1,n_2}{N,n}$ is nonzero only if $n_1+n_2=N+n$ since $\ket{n_1,n_2}$ and $\ket{N,n}$ are orthogonal when they do not belong to the same energy subspace. To order $z^2$, one can use the zero-temperature limit of the $T$ matrix but one must shift its argument by $2\mu$ to take into account the nonzero chemical potential. In the following we include the zero-point energy $\w_z/2$ in the chemical potential so that the quantized levels of the harmonic oscillator with frequency $\w_z$ are given by $\eps_n=n\w_z$ ($n\in\mathbb N$). We thus obtain  
	\begin{align} 
		\delta \bar n_\k &= \sum_{n_1,n_2\atop N,n} |U_{n_1,n_2;N,n}|^2 |\phi_n^-(0)|^2 \frac{2}{\beta^2} \sum_{\wn,\wnu} \sum_\K G(\k,n_1,i\wn)^2 G(\K-\k,n_2,i\wnu-i\wn)  t(i\wnu-\eps^+_{\K,N}+2\mu) \nonumber\\ 
		&= - \sum_{n_1,n_2\atop N,n} |U_{n_1,n_2;N,n}|^2 |\phi_n^-(0)|^2 \frac{2}{\beta} \sum_{\wnu} \sum_\K \frac{t(i\wnu-\eps^+_{\K,N}+2\mu)}{(i\wnu-\xi_\k-\xi_{\K-\k}-(N+n)\w_z)^2} \nonumber\\ 
		&= - \sum_{N,n} |\phi_n^-(0)|^2 \frac{2}{\beta} \sum_{\wnu} \sum_\K \frac{t(i\wnu-\eps^+_{\K,N}+2\mu)}{(i\wnu-\xi_\k-\xi_{\K-\k}-(N+n)\w_z)^2} ,
	\end{align}
	where we have kept only terms of order $z^2$ and used $\sum_{n_1,n_2}|U_{n_1,n_2;N,n}|^2=1$. Performing the sum over $\wnu$ by contour  integral, we finally obtain 
	\begin{align} 
		\delta \bar n_\k &= -2 \sum_{N,n} |\phi_n^-(0)|^2 \sum_\K \int_0^\infty \frac{d\w}{\pi} n_B(\w+\eps^+_{\K,N}-2\mu) \Im \llbrace \frac{t^R(\w)}{(\w+i0^+ +\eps_\K/2-\eps_\k - \eps_{\K-\k} - n\w_z)^2} \rrbrace \nonumber\\ 
		&= -2 z^2 \sum_{N,n} |\phi_n^-(0)|^2 \sum_{\K} e^{-\beta\eps^+_{\K,N}} \int_0^\infty \frac{d\w}{\pi} e^{-\beta \w}  \Im \llbrace \frac{t^R(\w)}{(\w+i0^+ +\eps_\K/2-\eps_\k - \eps_{\K-\k} - n\w_z)^2} \rrbrace 
		\label{app21} 
	\end{align}
	to order $z^2$. The contribution of the interactions to the pressure to the same order is obtained by considering the 2D density $\delta\bar n= \frac{1}{L^2} \int_\k \delta\bar n_\k=\partial \delta P/\partial\mu$, i.e.
	\begin{align}
		\delta P &= -\frac{z^2}{\beta}   \sum_{N,n} |\phi_n^-(0)|^2 \int_{\K,\k}
		e^{-\beta\eps^+_{\K,N}} \int_0^\infty \frac{d\w}{\pi} e^{-\beta \w}  \Im \llbrace \frac{t^R(\w)}{(\w+i0^+ -2\eps_\k -  n\w_z)^2} \rrbrace \nonumber\\ 
		&= z^2\sum_N \int_\K e^{-\beta\eps^+_{\K,N}} \int_0^\infty \frac{d\w}{\pi} e^{-\beta \w} \Im[\ln t^R(\w)] , 
		\label{app14}
	\end{align}
	where we have integrated by part in the integral over $\w$ and used~(\ref{app12}). 
	This results is in agreement with a similar virial calculation done for a 2D Fermi gas in \cite{Ngampruetikorn2013_sm}. It is however different from that of \cite{Ren2004_sm} for a two dimensional Bose gas, which we ascribe to the fact that that calculation is not truly a virial expansion but a perturbative expansion with ressumation of some of the $T$-matrix diagrams.
	
	Equation~(\ref{app14}) yields the contact 
	\begin{align} 
		\frac{C}{L^2} &= -2m^2z^2 \sum_N \int_\K  e^{-\beta\eps^+_{\K,N}} \int_0^\infty \frac{d\w}{\pi} e^{-\beta \w} \Im[t^R(\w)] \nonumber\\ 
		&= 2m^2z^2 \sum_N \int_\K  e^{-\beta\eps^+_{\K,N}} \sum_n |\phi^-_n(0)|^2  \int_\k e^{-\beta\eps^-_{\k,n}} |t^R(\eps^-_{\k,n})|^2 , 
		\label{app18}
	\end{align}
	where we have used~(\ref{app13}) and $\Im[t^R(\w)^{-1}]=\pi\sum_n |\phi_n^-(0)|^2 \int_\k  \delta(\w-\eps^-_{\k,n})$. 
	
	In the low-temperature regime $T\ll\w_z$, we can restrict ourselves to $N=n=0$ in~(\ref{app18}). Performing the sum over $\K$ and using~(\ref{app23}), one finds 
	\beq 
	\frac{C}{L^2} = 4(2\pi)^{3/2} z^2 \frac{\ell_z}{\lambda^2} \int_0^\infty dk \, k 
	\frac{e^{-\beta k^2/m}}{\left[\ln\left(\frac{ka_2}{2}e^\gamma\right)\right]^2 + \frac{\pi^2}{4}}  .
	\label{app18a}
	\eeq  
	Performing the integral in the limit $|\ln(\half\sqrt{ma_2^2T}e^\gamma)|\gg 1$~\cite{not22_sm}, we finally obtain 
	\beq 
	\frac{C_{\rm virial}}{L^2} = 2 (2\pi)^{5/2} \frac{\bar n^2\ell_z}{\left|\ln\left(\frac{a_2}{\lambda}\sqrt{\frac{2}{\pi}}e^\gamma\right)\right|^2} ,
	\eeq 
	where $\bar n=z/\lambda^2+\calO(z^2)$. The result~(\ref{app18a}) can also be obtained from Eq.~(8) in the main text, relating the contact to the pressure and the energy density, using the expression~(\ref{app14}) of the pressure.

	\subsection{C. Calculation of the contact from the pair distribution function}

	\subsubsection{1. Scattering states}
	
	The scattering states associated with the relative motion can be deduced from the Lippmann-Schwinger equation. They satisfy 
	\beq 
	\ket{\psi_{\k,n}} = \ket{\k,n} + \frac{1}{L^2} \sum_{\k',n'} \frac{t^R_{n'n}(\eps^-_{\k,n})}{\eps^-_{\k,n}+i0^+ - \eps^-_{\k',n'}} \ket{\k',n'} ,
	\eeq 
	where $t^R(\w)=t(\w+i0^+)$ is the retarded $T$ matrix defined by~(\ref{app11}).  Using 
	\beq
	\int_{\k'} \frac{e^{i\k'\cdot\rhobf}}{\eps^-_{\k,n}+i0^+ - \eps^-_{\k',n'}} = -i \frac{m}{4} H_0^{(1)}(k_{n'}\rho) , 
	\label{app15} 
	\eeq 
	where $k_{n'}^2=\k^2+(n-n')m\w_z$, $\rho=|\rhobf|$ and $H_0^{(1)}$ is a Hankel function of the first kind, we obtain 
	\beq 
	\psi_{\k,n}(\r) = \frac{e^{i\k\cdot\rhobf}}{L} \phi^-_{n}(z) - i \frac{m}{4L} \sum_{n'} t^R(\eps^-_{\k,n}) \phi^-_{n'}(0)^* \phi^-_{n}(0) \phi^-_{n'}(z)  H_0^{(1)}(k_{n'}\rho) . 
	\label{app16} 
	\eeq
	When $k_{n'}^2<0$, $k_{n'}$ in~(\ref{app15}) and (\ref{app16}) must be understood as $i|k_{n'}|$ and $H_0^{(1)}(i|k_{n'}|\rho) = -(2i/\pi) K_0(|k_{n'}|\rho)$ decays exponentially for large $\rho$, which corresponds to a closed scattering channel. 
	
	\subsubsection{2. Pair distribution function}
	
	We now consider the pair distribution function, averaged over the position $\R=(\r_1+\r_2)/2$ of the pair center of mass, 
	\beq 
	g(\r_1-\r_2) = \int d^3 R \, \mean{\hat\psi^\dagger(\r_1)\hat\psi^\dagger(\r_2)  \hat\psi(\r_2)\hat\psi(\r_1)} .
	\eeq 
	To obtain the contact to order $z^2$, it is sufficient to consider states with only two particles. The contribution to the pair distribution function of the two-particle scattering state $\ket{\psi^{(s)}_{\K,N;\k,n}}$, defined by its properly symmetrized wavefunction
	\beq 
	\psi^{(s)}_{\K,N;\k,n}(\R,\r) = \frac{1}{\sqrt{2}} \frac{e^{i\K\cdot\R_\para}}{L} \phi^+_N(Z) [ \psi_{\k,n}(\r) + \psi_{-\k,n}(\r) ] ,
	\eeq 
	reads $\int d^3R\, 2|\psi^{(s)}_{\K,N;\k,n}(\R,\r)|^2$. To order $z^2$, the function $g(\r)$ is obtained by summing over all states $\psi^{(s)}_{\K,N;\k,n}$ with the Boltzmann weight 
	\beq 
	\frac{1}{Z} e^{-\beta(\eps^+_{\K,N}+ \eps^-_{\k,n} -2\mu)} = z^2 e^{-\beta(\eps^+_{\K,N}+ \eps^-_{\k,n})} + \calO(z^3) 
	\eeq 
	(using $Z=1+\calO(z)$), i.e. 
	\beq
	g(\r) = \half \sum_{\K,\k} \sum_{N,n} z^2 e^{-\beta(\eps^+_{\K,N}+ \eps^-_{\k,n})} \int d^3R\,   2 |\psi^{(s)}_{\K,N;\k,n}(\R,\r)|^2 ,
	\label{app17} 
	\eeq 
	where a factor $1/2$ has been introduced in the sum over $\k$ since $\psi^{(s)}_{\K,N;\k,n}$ and $\psi^{(s)}_{\K,N;-\k,n}$ are the same state. 
	
	Let us first consider the short distance limit $r=|\r|=\sqrt{\rho^2+z^2}\ll \lambda,\ell_z$. In that case one has~\cite{Petrov00a_sm} 
	\beq 
	\frac{i}{4} \sum_{n'} \phi^-_{n'}(z)  \phi^-_{n'}(0)^* H_0^{(1)}(k_{n'}\rho) \simeq  \frac{1}{4\pi}\left(\frac1r-\frac1{a_3} \right)+\frac1{m t^R(\eps^-_{\k,n})}
	\label{app20b}
	\eeq 
	and thus
	\beq 
	\psi_{\k,n}(\r) \simeq - \frac{m}{4\pi L}\phi_n^-(0) t^R(\eps^-_{\k,n})\left(\frac1r-\frac1{a_3}\right) .
	\label{app20c}
	\eeq 
	Equation~(\ref{app17}) then gives 
	\beq 
	g(\r) = \frac{2m^2z^2}{(4\pi rL)^2} \sum_{\K,N} e^{-\beta\eps^+_{\K,N}}  \sum_{\k,n} e^{-\beta\eps^-_{\k,n}} |\phi_n^-(0)|^2|t^R(\eps^-_{\k,n})|^2  = \frac{C}{(4\pi)^2} \left( \frac{1}{r} - \frac{1}{a_3} \right)^2 ,
	\label{app20} 
	\eeq
	where $C$ is the contact~(\ref{app18}). At short length scales, $r\ll\lambda,\ell_z$, the pair distribution function behaves as in an isotropic 3D system with contact $C$. 
	
	Let us now consider the intermediate length scales, $\ell_z\ll\rho\ll\lambda$, in the low-temperature regime $T\ll\w_z$. Since only the states with $\eps^-_{\K,N},\eps^-_{\k,n}\lesssim T$ contribute significantly to the pair distribution function~(\ref{app17}), we can restrict ourselves to $N=n=0$ and thus consider only the scattering states 
	\beq 
	\psi_{\k,0}(\r) = \frac{e^{i\k\cdot\rhobf}}{L} \phi^-_{0}(z) - i \frac{m}{4L} \sum_{n'} t^R(\eps^-_{\k}) \phi^-_{n'}(0)^* \phi^-_{0}(0) \phi^-_{n'}(z)  H_0^{(1)}(k_{n'}\rho) 
	\label{app19} 
	\eeq  
	with $|\k|\lesssim 1/\lambda$. If $n'>0$ then $k_{n'}^2<0$ and $H_0^{(1)}(k_{n'}\rho)=H_0^{(1)}(i|k_{n'}|\rho)\sim e^{-\sqrt{n'}\rho/\ell_z}$ decays exponentially for $\rho\gg\ell_z$. On the other hand, if $n'=0$, 
	\beq
	H_0^{(1)}(k_{n'}\rho)=H_0^{(1)}(|\k|\rho)\simeq \frac{2i}\pi\left(\ln\frac{|\k|\rho}2+\gamma-i\frac\pi2\right)
	\eeq
	for $\rho\ll\lambda$ (since $|\k|\lesssim 1/\lambda$). Thus for $\ell_z\ll\rho\ll\lambda$, we can just consider the contribution of $n'=0$ in~(\ref{app19}) and using~\eqref{app23} we obtain
	\beq 
	\psi_{\k,0}(\r) \simeq \frac{m}{2\pi L}  t^R(\eps^-_{\k})|\phi^-_{0}(0)|^2 \phi^-_{0}(z)  \ln(\rho/a_2) .
	\label{app19bis}
	\eeq  
	From~(\ref{app17}) we finally deduce  
	\begin{align}
		g(\r) &= \frac{2m^2z^2}{L^2} \left( \frac{\ln(\rho/a_2)}{2\pi}\right)^2 \sum_{\K,\k} e^{-\beta(\eps^+_{\K}+ \eps^-_{\k})} |t^R(\eps^-_{\k})|^2 |\phi^-_0(0)|^4 |\phi^-_0(z)|^2 \nonumber\\ 
		&= \left( \frac{\ln(\rho/a_2)}{2\pi}\right)^2 C |\phi^-_0(0)|^2 |\phi^-_0(z)|^2 \nonumber\\ 
		&=  \left( \frac{\ln(\rho/a_2)}{2\pi}\right)^2 \frac{C}{\sqrt{2\pi}\ell_z}  |\phi^-_0(z)|^2 ,
		\label{app20a} 
	\end{align}
	where $C$ is the low-temperature limit of the contact~(\ref{app18}). If we average over $z$, we recover the standard result for a 2D system with a contact $C/\sqrt{2\pi}\ell_z$. The latter is related to the pressure by 
	\beq 
	\frac{C}{\sqrt{2\pi}\ell_z L^2} = 4 \sqrt{2\pi} \frac{m}{\ell_z} \frac{\partial P}{\partial (1/a_3)} = -4\pi m \frac{\partial P}{\partial \ln a_2}  
	\eeq 
	and therefore coincides with the usual thermodynamic definition of a 2D contact. 
	
	We shall see in Sec.~D that the short-distance behavior of the pair distribution function, Eqs.~(\ref{app20}) and (\ref{app20a}), is actually valid only when $r$ and $\rho$ are much smaller than the interparticle distance $d$.

	\subsubsection{3. Momentum distribution}
	
	The knowledge of the short-distance behavior of the pair distribution function allows one to obtain the high-momentum behavior of the momentum distribution function $n_{\k}$. The calculation is standard and yields 
	\beq 
	\bar n_{\k,k_z} = \frac{C}{(\k^2+k_z^2)^2} \quad \mbox{if} \quad \frac{1}{d},\frac{1}{\lambda},\frac{1}{a_3}\ll \sqrt{\k^2+k_z^2} ,
	\label{app22}
	\eeq 
	and 
	\beq 
	\bar n_\k = \int \frac{dk_z}{2\pi} \bar n_{\k,k_z} = \frac{1}{\sqrt{2\pi}\ell_z} \frac{C}{|\k|^4} \quad \mbox{if} \quad  \frac{1}{d},\frac{1}{\lambda} \ll |\k| \ll \frac{1}{\ell_z} \quad \mbox{and} \quad T\ll \w_z ,
	\label{app22a} 
	\eeq 
	where we assume the normalization $\frac{1}{L^2}\sum_{\k} \int_{k_z} \bar n_{\k,k_z}=N$ (with $N\equiv\mean{\hat N}$ the total number of particles).
	Equation~(\ref{app22a}) can also be easily obtained from~(\ref{app21}). Again we find that the 3D contact~(\ref{app24}) governs the high-momentum behavior but the intermediate range $1/d,1/\lambda\ll|\k|\ll 1/\ell_z$ involves the 2D contact $C/\sqrt{2\pi}\ell_z$.

	\subsection{D. Beyond the virial expansion}  
	
	Results similar to~(\ref{app20},\ref{app20a}) and (\ref{app22},\ref{app22a}) have been obtained in Refs.~\cite{He19_sm,Bougas20_sm,Decamp2018_sm} for fermions and bosons in quasi-1D and quasi-2D traps. They are thus valid beyond the virial expansion as can be easily shown following the method of Werner and Castin~\cite{Werner12a_sm,Werner12_sm}. 
	We sketch the proof here.
	
	Standard contact theory assumes that when two particles are close to each other ($b\ll r_{ij}\ll d$ with $b$ the range of the interaction potential, $d$ the typical interparticle distance, and $r_{ij}=|\r_{ij}|= |\r_i-\r_j|$), the many-body wavefunctions take the asymptotic form
	\begin{equation}
		\Psi(\r_1,\ldots,\r_N)\simeq \psi^{(s)}_0(\r_{ij})\calA_{ij}(\R_{ij},(\r_k)_{k\neq i,j}),
	\end{equation}
	where $\psi^{(s)}_0(\r_{ij})$ is the zero-energy scattering state. Since the range $b$ of the interaction potential and the 3D $s$-wave scattering length $a_3$ are much smaller than the harmonic oscillator length $\ell_z$, $\psi^{(s)}_0(\r)$  varies as $1/r-1/a_3$ when $b\ll r\ll \ell_z,d$,  as for an isotropic 3D system. Normalizing $\psi^{(s)}_0(\r)$ such that $\psi^{(s)}_0(\r)\simeq 1/r-1/a_3$  in that limit, the contact of the state $\Psi$ is then given by
	\begin{equation}
		C_\Psi= 32\pi^2 \sum_{i<j}\int\prod_{k\neq i,j} d^3r_k\int d\R_{ij}|\calA_{ij}(\R_{ij},(\r_k)_{k\neq i,j}|^2,
	\end{equation}
	which, thanks to the normalization of $\psi^{(s)}_0$, is related to the energy $E_\Psi$ of the state  $\Psi$ by $C_\Psi=8\pi m\frac{\partial E_\Psi}{\partial(1/a_3)}$, and the pair distribution function behaves as $g_{\Psi}(\r)= [C_\Psi/(4\pi)^2](1/r-1/a_3)^2$  for $b\ll r\ll \ell_z,d$. Averaging over the states with a Boltzmann weight, we recover the thermodynamic contact discussed in the main text.
	
	In the quasi-2D case, there are two points to consider. First, the zero-energy limit must be taken carefully (since the $T$ matrix vanishes at zero energy). Second, when $\ell_z\ll d$, the short distance behavior will cross-over from a 3D to a 2D behavior depending on whether $r_{ij}\ll \ell_z$ or $r_{ij}\gg \ell_z$ (in the opposite regime, $\ell_z\gg d$, only the 3D behavior is visible at short distances $r_{ij}\ll d$).
	
	Focusing on the quasi-2D regime {$T\ll\w_z$} in the harmonic trap, the short-distance behavior of $\psi^{(s)}_0(\r)$ is obtained from that of $\psi_{\k,0}(\r)$ in the limit $|\k|\to 0$, see Eq.~\eqref{app19}. In the short-distance limit, the symmetrization only changes the wavefunction normalization by a factor $\sqrt{2}$. As this factor is unimportant here (it will be absorbed in $\calA$ and is the same in the 3D and 2D scattering regimes), we  discard it. 
	
	In the 3D scattering regime, we can use Eq.~\eqref{app20c} with $n=0$ and $|\k|\to 0$.
	Therefore, to obtain a correctly normalized $\psi^{(s)}_0(\r)$, we need to normalize $\psi_{\k,0}(\r)$  by a factor $\mathcal N=-\frac{m}{4\pi L} \phi^-_0(0) t^R(\eps^-_{\k,0})$. We note that this normalization vanishes at low energies,  as expected for a system which is effectively 2D, see Appendix~A in~\cite{Werner12a_sm}. 
	
	In the 2D scattering regime, $\rho\gg \ell_z$, we can use Eq.~\eqref{app19bis}, and the properly normalized scattering state (using the factor $\mathcal N$ computed above) reads 
	\begin{equation}
		\psi^{(s)}_0(\r) \simeq -2\phi^-_0(z)\phi^-_0(0)^*\left[\log(\rho/a_2)+\ldots\right] 
	\end{equation}
	and is finite at zero energy.
	To summarize, we find that when two particles are close to each other, the many-body wavefunction behaves as
	\begin{equation}
		\Psi(\r_1,\ldots,\r_N)=\calA_{ij}(\R_{ij},(\r_k)_{k\neq i,j})\times
		\begin{cases}
			\left(\frac{1}{r_{ij}}-\frac{1}{a_3}\right) & \text{if } b\ll r_{ij}\ll \ell_z,d,\\
			-2\phi^-_0(0)^*\phi^-_0(z_{ij})\log(\rho_{ij}/a_2)              & \text{if }  \ell_z\ll \rho_{ij}\ll d ,
		\end{cases}
	\end{equation}
	with $z_{ij}=z_i-z_j$ and $\rho_{ij}=|\rhobf_i-\rhobf_j|$. Since $|\phi_0^-(0)|^2=1/\sqrt{2\pi}\ell_z$, this implies that the pair  distribution function of an arbitrary many-body state $\Psi$ behaves as
	\begin{equation}
		g_{\Psi}(\r)= 
		\begin{cases}
			\frac{C_\Psi}{(4\pi)^2}\left( \frac{1}{r}-\frac{1}{a_3} \right)^2  & \text{if } b\ll r\ll \ell_z,d,\\
			\frac{C_\Psi}{\sqrt{2\pi}\ell_z}|\phi^-_0(z)|^2 \left(\frac{\ln(\rho/a_2)}{2\pi}\right)^2        & \text{if } \ell_z\ll \rho\ll d  .
		\end{cases}
	\end{equation}
	By integrating over $z$ we recover the usual expression of the 2D case with a contact $C_\Psi/\sqrt{2\pi}\ell_z$. This generalizes the conclusion drawn from the virial expansion in Sec.~II.C.

	\section{III.\;\; Contact and classical field simulations}  
	
	
	\begin{figure}
		\centering
		\includegraphics[width=8cm]{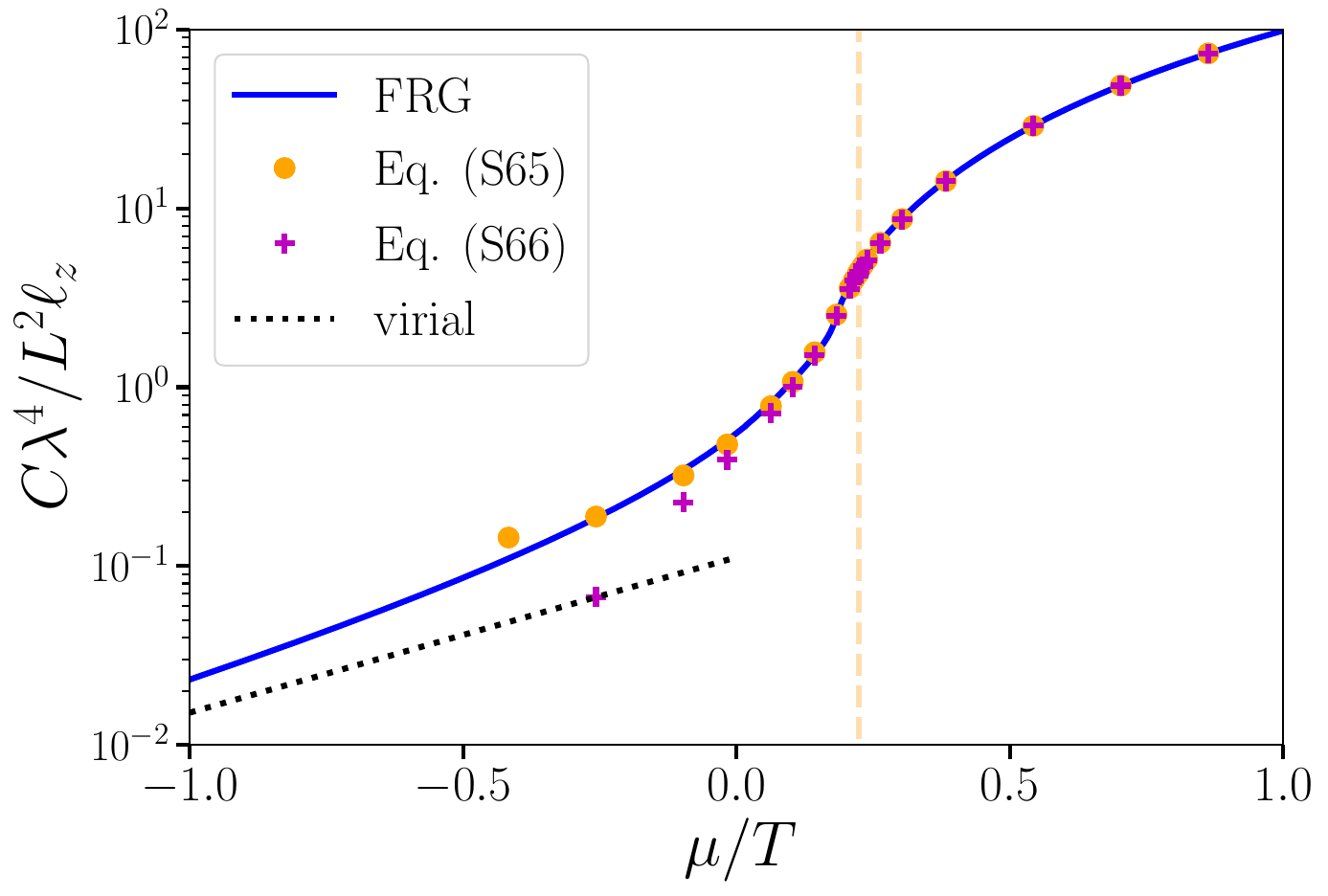}
		\caption{Reproduction of the inset of Fig.~1 in the main text: normalized contact $\cquasi\lamb^4/L^2\ell_z$ {\it vs} $\mu/T$ for $2mg=0.32$ and $T/\w_z\simeq 0.00145$ as obtained from the FRG (solid (blue) line) and classical field simulations using the thermodynamic definition~\eqref{eq_MC_th} (orange circles) and the expectation value $\mean{|\varphi(0)|^4}$ [Eq.~\eqref{eq_MC_phi4}] (purple crosses). The vertical dashed line shows the position of the BKT transition as estimated from \eqref{eq_muc}.  }
		\label{fig_CSM}
	\end{figure}

	The contact of the planar Bose gas can be obtained from the classical field simulations (CFS) of \cite{Prokofev02_sm} using the thermodynamic definition \eqref{app24}, following the method of \cite{Ozawa2014_sm,Ota2018_sm}.
	
	Ref.~\cite{Prokofev02_sm} gives the density  near the BKT transition as
	\begin{equation}
		\bar n_{\rm CFS} = \bar n_c + F\left(X\right),
	\end{equation}
	where $\lambda^2 F$ is a universal function, $X=(\mu-\mu_c)/mg T$ measures the distance from the transition point at $\mu=\mu_c$ (neglecting the renormalization of the interaction in the weak-interaction regime) and
	\begin{equation}
		\begin{split}
			\bar n_c &= \frac{mT}{2\pi}\log\left(\frac{\xi}{mg}\right),\\
			\mu_c &= \frac{mgT}{\pi}\log\left(\frac{\xi_\mu}{mg}\right),
		\end{split}
		\label{eq_muc}
	\end{equation}
	with $\xi\simeq 380$ and $\xi_\mu\simeq 13.2$. Integrating the density with respect to the chemical potential numerically, we obtain the pressure up to a constant,
	\begin{equation}
		\Delta P_{\rm CFS} = P_{\rm CFS}-P_c = mg T \biggl(X\bar n_c+\int_0^X dx\, F\left(x\right)\biggr).
	\end{equation}
	Finally, using the thermodynamic definition~\eqref{app24} of the contact, i.e. 
	\beq 
	\frac{C}{L^2} = -2 \sqrt{2\pi} \ell_z (mg)^2 \frac{\partial P}{\partial g} , 
	\eeq 
	we obtain 
	\begin{equation}
		\begin{split}
			\frac{C_{{\rm CFS},1}}{L^2} = \frac{C_c}{L^2}+2\sqrt{2\pi}\ell_z m^2g \left[ (\mu-\mu_c)(\bar n_{\rm CFS}+\lambda^{-2})+\left(\mu_c-\frac{mg T}{\pi}\right)(\bar n_{\rm CFS}-\bar n_c)-\Delta P_{\rm CFS}\right] .
		\end{split}
		\label{eq_MC_th}
	\end{equation}
	The constant $C_c$, corresponding to the contact at the BKT transition, is fixed by minimizing the relative difference between FRG and classical field theory calculations. Figure~\ref{fig_CSM}, left panel, shows the FRG and CFS results (orange symbols) for the same parameters as used in the main text, $2mg=0.32$ and $T/\omega_z\simeq 0.00145$. We find a very good agreement between the two methods, once $C_c$ is  fixed, except in the very-low-density limit.  
	
	The contact can also be estimated from the expectation value $\langle |\varphi(0)|^4\rangle$ where $\varphi(\rhobf)$ is the classical field. Indeed, the relation $\frac{\partial P_{\rm CFS}}{\partial g}= -\frac12 \int d^2\rho\, \langle |\varphi(\rhobf)|^4\rangle$, implies
	\begin{equation}
		\begin{split}
			\frac{C_{{\rm CFS},2}}{L^2} = \sqrt{2\pi} \ell_z (mg)^2 \langle |\varphi(0)|^4\rangle.
		\end{split}
		\label{eq_MC_phi4}
	\end{equation}
	The expectation value $\langle |\varphi(0)|^4\rangle$ can be deduced from the function $Q=2\langle |\varphi(0)|^2\rangle^2-\langle |\varphi(0)|^4\rangle$ given in~\cite{Prokofev02_sm}. The corresponding contact is shown as purple crosses in Fig.~\ref{fig_CSM}. The two classical field contacts, given by~(\ref{eq_MC_th}) and (\ref{eq_MC_phi4}), are in good agreement at positive chemical potential, but the estimate~\eqref{eq_MC_phi4} deteriorates when $\mu<0$ (in fact it even becomes negative for $\mu/T\simeq-0.41$). Since $\langle |\varphi(0)|^4\rangle\simeq 2\langle |\varphi(0)|^2\rangle$ and $\langle |\varphi(0)|^2\rangle$ becomes exponentially small in that limit, the failure of~\eqref{eq_MC_phi4} might come from the subtraction of two small numbers. The estimate~(\ref{eq_MC_th}) also eventually fails in the low-density limit since it gives $C/C_{0}\simeq 4.8$  for the smallest available value of the density ($\mu/T\simeq-0.41$), whereas one expects $C/C_{0}\simeq 2$ from the virial expansion.
	

	We compare the contact $C_{{\rm CFS},1}$ with $C_{\rm FRG}$ in Fig.~\ref{fig:CMC_vs_CFRG}. When they are considered as a function of the chemical potential $\mu$ (blue points in the figure), they nicely agree in the large-$\mu$ limit. However, when they are considered as a function of the  phase-space density $\calD$ (red points in the figure), this is not the case anymore as one can see a systematic shift between $C_{{\rm CFS},1}$ and $C_{\rm FRG}$ even in the large-density limit. The disagreement is more pronounced at $T=0.03\w_z$ than $T=0.00145\w_z$.  The difference between FRG and CFS comes from the estimation of the density. While the FRG and CFS densities differ only by $5\%$ (for a given value of the chemical potential), this is enough to have a visible effect. Note that the difference in the estimation of the density changes not only the normalization of the contact (and hence the position on the $y$-axis in Fig.~2 of the main text) but also the value of the phase-space density (and hence the position on the $x$-axis).

	\begin{figure}
		\centering
		\begin{subfigure}
			\centering
			\includegraphics[width=8cm]{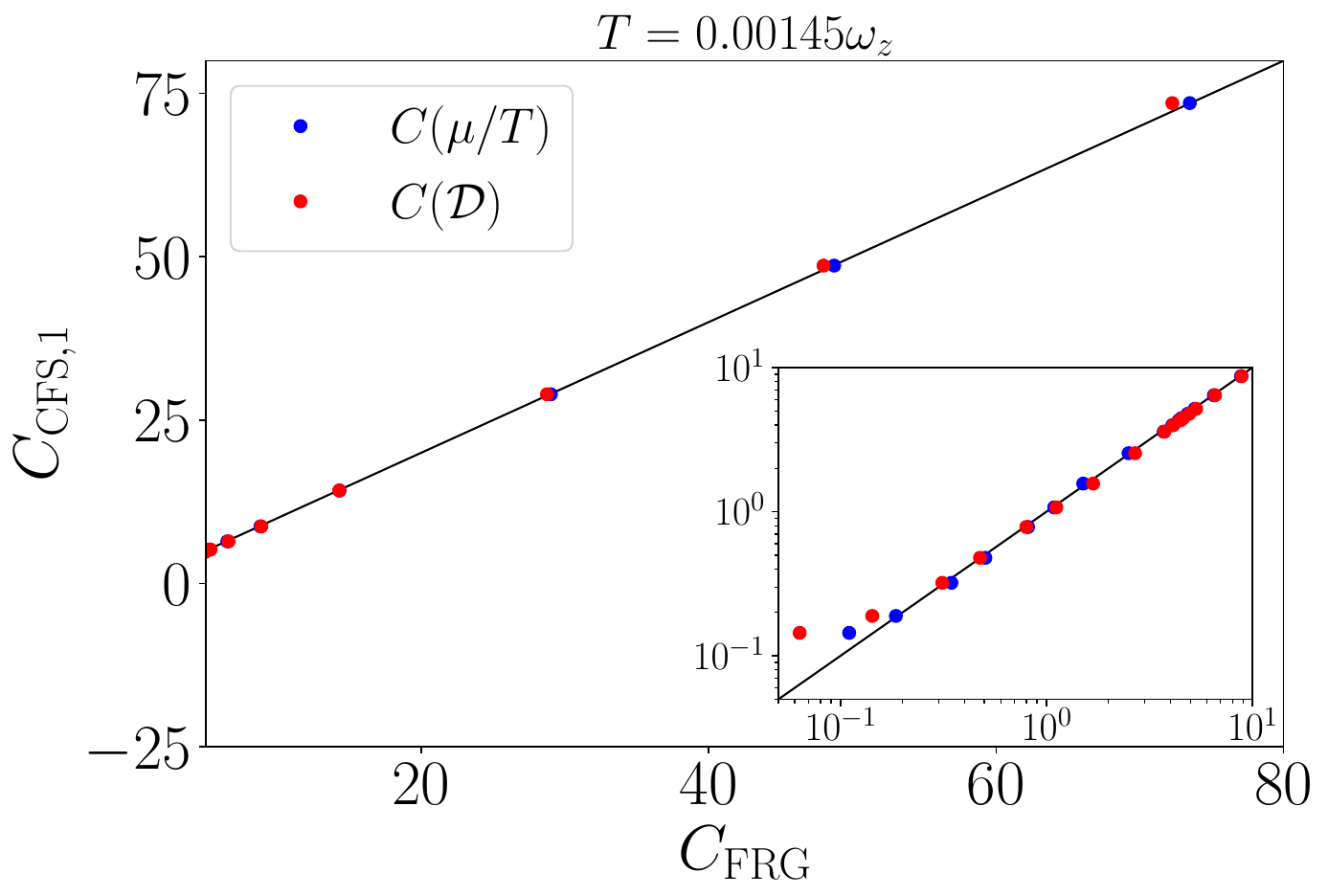}
		\end{subfigure}%
		\begin{subfigure}
			\centering
			\includegraphics[width=8cm]{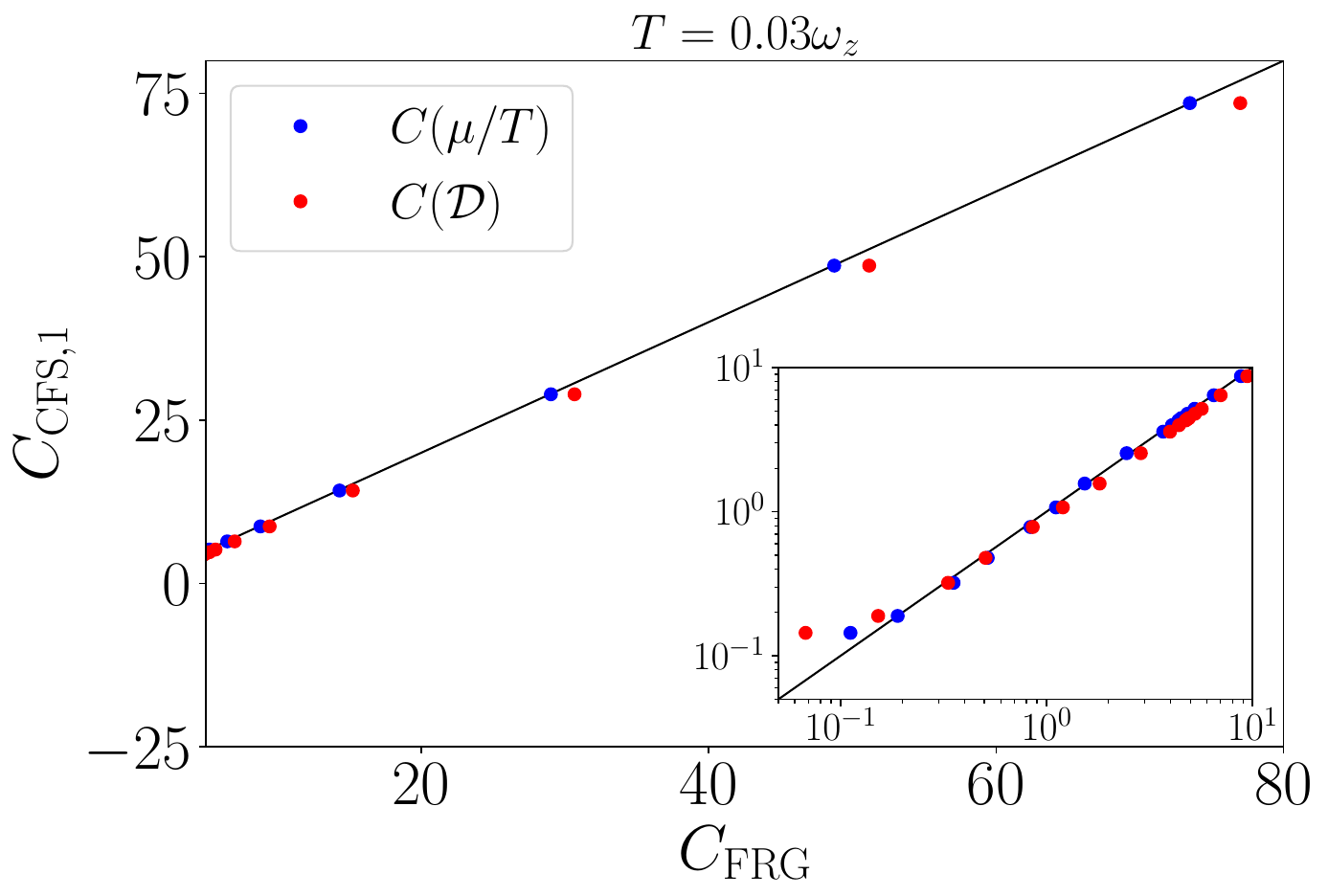}
		\end{subfigure}
		\caption{Scatter plot of the CFS and FRG contacts, as functions of $\mu/T$ (blue points) or phase-space density $\mathcal D$ (red points). The inset corresponds to a log-log scale at low contact. The thin black line corresponds to the curve $C_{{\rm CFS},1}=C_{\rm FRG}$. Left: $T=0.00145\omega_z$. Right:  $T=0.03\omega_z$, as used in Fig.~2 of the main text for comparison to the experimental data.}
		\label{fig:CMC_vs_CFRG}
	\end{figure}


	\section{IV.\;\; Normalization of the contact}  
	
	The contact varies roughly as the square of the density and the latter changes by several orders of magnitude between the low-density and superfluid regimes. It is therefore convenient to normalize the contact using the density. In Ref.~[14] and in the main text, the mean-field contact $C_0/L^2 = 4(2\pi)^{3/2}\bar n^2 a_3^2/\ell_z$ is used for this purpose. It could be argued that it is more natural to use the Bogoliubov contact~\eqref{CBog}, since $C/C_{\rm Bog}$ tends to one in the deep superfluid regime $\calD\gg 1$.
	
	\begin{figure}
		\centering
		\includegraphics[width=8cm]{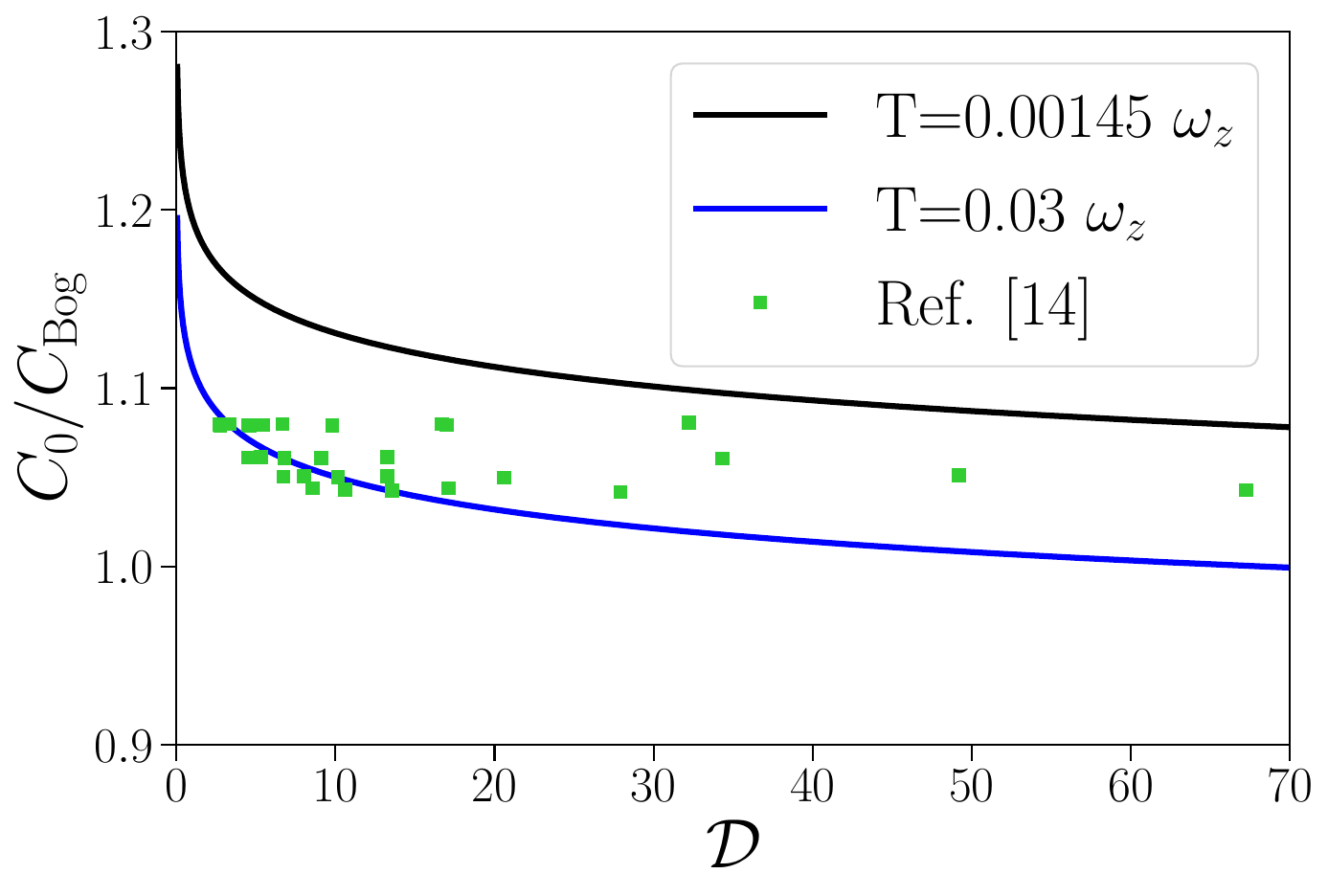}
		\caption{Ratio $C_0/C_{\rm Bog}$ between the mean-field contact and the Bogoliubov contact at two different temperatures. The corresponding ratio extracted from the experimental data of Ref.~[14] is shown by (green) symbols.}
		\label{fig:C0}
	\end{figure}

	\begin{figure}
		\centering
		\begin{subfigure}
			\centering
			\includegraphics[width=8cm]{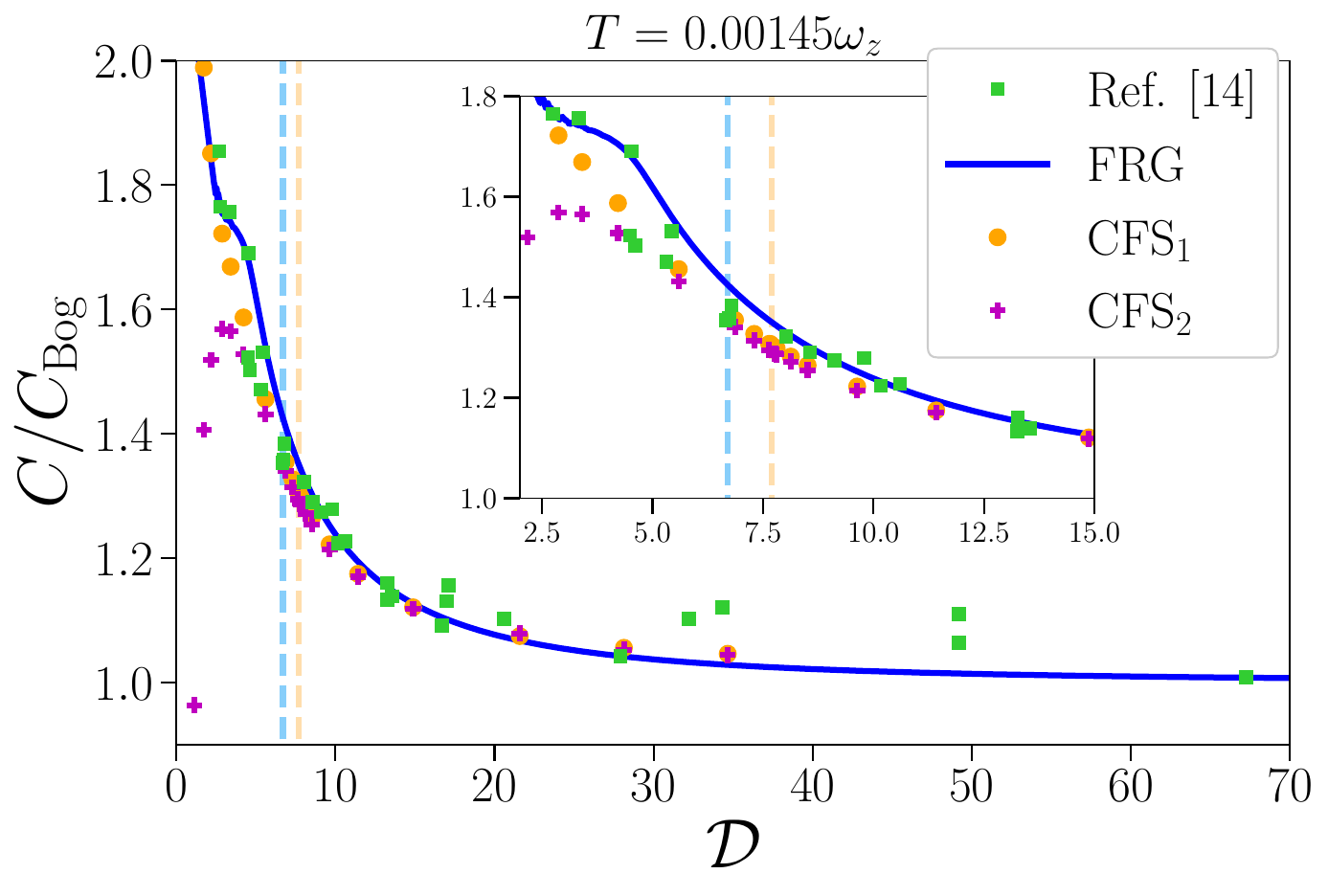}
		\end{subfigure}%
		\begin{subfigure}
			\centering
			\includegraphics[width=8cm]{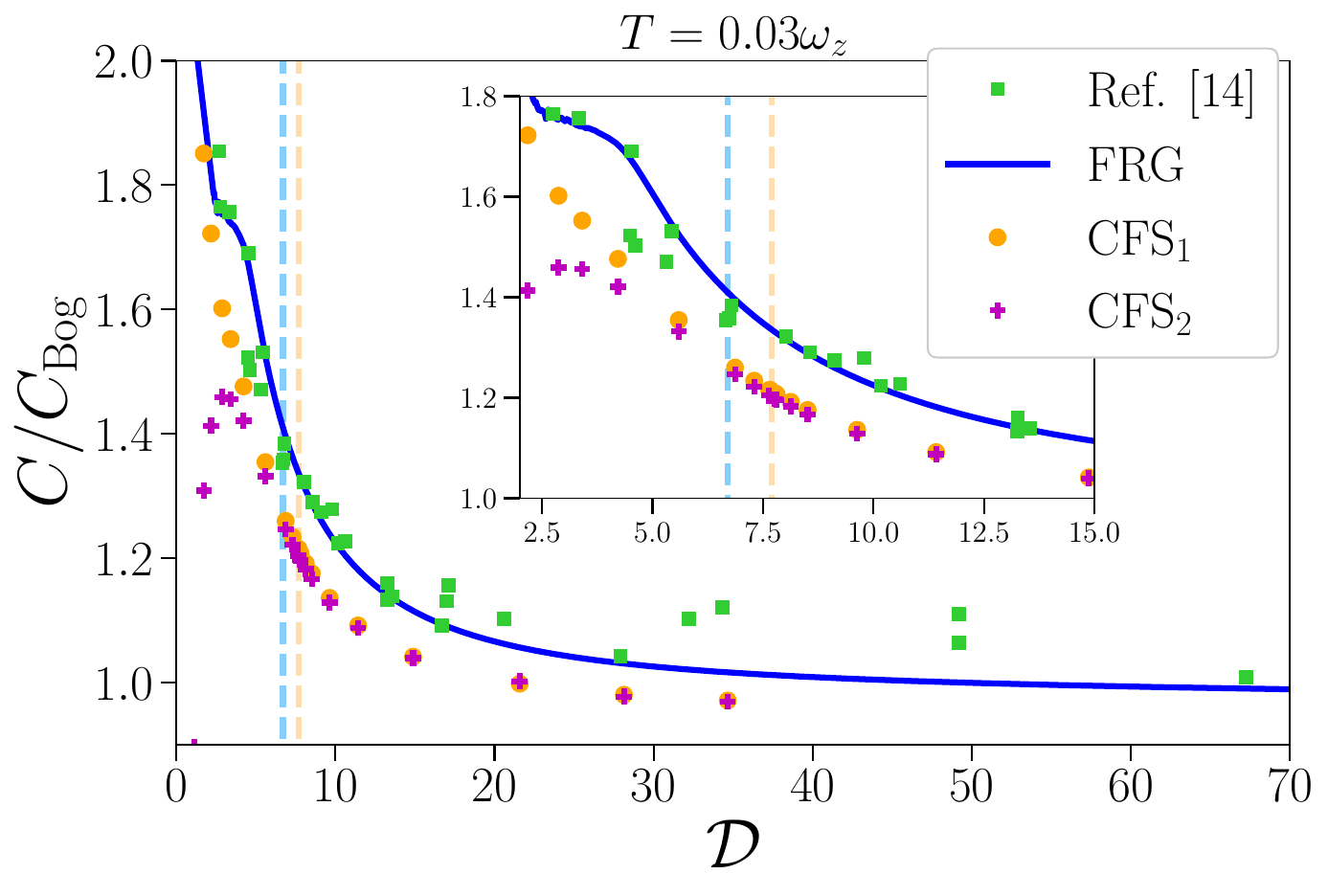}
		\end{subfigure}
		\caption{Normalized contact $C/C_{\rm Bog}$ using the Bogoliubov contact for $T=0.00145\omega_z$ (left) and  $T=0.03\omega_z$ (right) obtained from FRG and CFS. Also shown are the experimental data of Ref.~[14].}
		\label{fig:CBog}
	\end{figure}

	The ratio $C_0/C_{\rm Bog}$ is shown in Fig.~\ref{fig:C0} for $mg=0.16$ and two temperatures, $T/\omega_z=0.00145$ and $T/\omega_z=0.03$. At low temperatures, the logarithmic corrections (responsible for the difference between $C_{\rm Bog}$ and $C_0$) are stronger. We also show the corresponding ratio in the experiments of Ref.~[14]. 
	
	Normalizing the contact with $C_{\rm Bog}$,  instead of $C_0$ as in Fig.~2 of the main text, gives  Fig.~\ref{fig:CBog}. For $T=0.03\w_z$, the two figures are very similar since the ratio $C_0/C_{\rm Bog}$ is close to unity. For $T=0.00145\w_z$, the normalized CFS contact $C_{\rm CFS}/C_{\rm Bog}$ is close to both the FRG result (in agreement with the left panel of Fig.~\ref{fig:CMC_vs_CFRG}) and the experimental data.

\newpage

%


\begin{thebibliography}{44}
\bibitem [{not({\natexlab{a}})}]{not55}%
\BibitemOpen
\href@noop {} {} \bibinfo {note} {Note that mean-field
theory (with $\tilde g(T)=\tilde g=2mg$) violates the equivalence between the
definitions~(\ref{Cdef}) and (\ref{Cdef2}) of the contact. It gives a
vanishing contact using~(\ref{Cdef2}) but a nonzero one
from~(\ref{Cdef}).}\BibitemShut {Stop}%
\bibitem [{not({\natexlab{b}})}]{not33}%
\BibitemOpen
\href@noop {} {} \bibinfo {note} {The Lee-Huang-Yang
correction does not modify the expression of the contact as a function of
$\mu$ but modifies the relation between $\mu$ and the mean density $\bar
n$}\BibitemShut {NoStop}%
\bibitem [{not({\natexlab{b}})}]{not34}%
\BibitemOpen
\href@noop {} {} \bibinfo {note} {To leading order, the Bogoliubov result is deduced from~(\ref{Cbog}) using the relation between $\mu$ and $\bar n$~\cite{notSM,not33},
$C_{\rm Bog}/L^2 = (2\pi)^{5/2} \bar n^2\ell_z/ 
|\ln( a_2 \sqrt{\pi \bar n} e^{\gamma+1/2})|^2$, 
although in practice we invert numerically $\bar n(\mu)$ to compute $C_{\rm Bog}$}\BibitemShut {NoStop}%
\end{thebibliography}
\begin{thebibliography}{19}
\bibitem [{not()}]{not22_sm}%
\BibitemOpen
\href@noop {} {}\bibinfo {note} {We set $k=\sqrt{mT}e^y$ and use the
condition $|\ln(\half\sqrt{ma_2^2T}e^\gamma)|\gg 1$ to approximate
$e^{-e^{2y}}$ by $\Theta(-y)$ (with $\Theta(y)$ the step
function).}\BibitemShut {Stop}%
\end{thebibliography}

\end{document}